\documentclass[11pt,a4paper]{article}
\usepackage{jheppub}
\usepackage{ifpdf}
\usepackage{multicol,bbm}
\usepackage{graphicx,psfrag}
\usepackage{dcolumn}
\usepackage{bm}
\usepackage{amsmath,amssymb, bbold}
\usepackage{color}
\usepackage[noabbrev]{cleveref}
\usepackage{mathtools}
\allowdisplaybreaks[1]
\let\normalcolor\relax

\makeatletter
\def\@fpheader{Prepared for submission to JCAP}
\makeatother

\graphicspath{ {./figures/} }

\newcommand{\Ms}{M_\star}

\title{Linear growth of structure in massive bigravity}

\author[a]{Adam R. Solomon,}
\author[b]{Yashar Akrami,}
\author[c]{Tomi S. Koivisto}
\affiliation[a]{DAMTP, Centre for Mathematical Sciences, University of Cambridge\\
       Wilberforce Rd., Cambridge CB3 0WA, UK}
\affiliation[b]{Institute of Theoretical Astrophysics, University of Oslo\\
       P.O. Box 1029 Blindern, N-0315 Oslo, Norway}
\affiliation[c]{Nordita, KTH Royal Institute of Technology and Stockholm University\\
       Roslagstullsbacken 23, SE-10691 Stockholm, Sweden}

\emailAdd{a.r.solomon@damtp.cam.ac.uk}
\emailAdd{yashar.akrami@astro.uio.no}
\emailAdd{tomi.koivisto@nordita.org}

\abstract{The ghost-free theory of massive gravity with two dynamical metrics has been shown to produce viable cosmological expansion, where the late-time acceleration of the Universe is due to the finite range of the gravitational interaction rather than a nonzero cosmological constant. Here the cosmological perturbations are studied in this theory. The full perturbation equations are presented in a general gauge and analyzed, focusing on subhorizon scales in the quasistatic limit during the matter-dominated era. An evolution equation for the matter inhomogeneities and the parameters quantifying the deviations from general relativistic structure formation are expressed in terms of five functions whose forms are determined directly by the coupling parameters in the theory. The evolution equation has a similar structure to Horndeski-type scalar-tensor theories, exhibiting a modified growth rate and scale-dependence at intermediate wavenumbers. Predictions of the theory are confronted with observational data on both background expansion and large-scale structure, although care must be taken to ensure a model is stable. It is found that while the stable models fit the data well, they feature deviations from the standard cosmology that could be detected or ruled out by near-future experiments.}

\preprint{NORDITA-2014-46}
\keywords{modified gravity, bigravity, massive gravity, cosmological perturbations, structure formation, cosmic acceleration, dark energy}

\begin{document}

\maketitle

\section{Introduction}

There has recently been considerable progress in the development of massive gravity and, as a direct result, bimetric gravity. To give the graviton a mass in a covariant local field theory, one needs to introduce a reference metric. It is then natural to give this second metric its own dynamics.

The linear theory of massive gravity has been known since the seminal paper of Fierz and Pauli~\cite{Fierz_et_Pauli1939}, and studies of interacting spin-2 field theories also have a long history~\cite{Isham:1971gm}. However, there had long been an obstacle to the construction of a fully nonlinear theory of massive gravity in the form of the notorious Boulware-Deser ghost~\cite{Boulware_et_Deser1939}, a pathological mode that propagates in massive theories at the nonlinear order. This ghost mode was thought to be fatal to massive and interacting bimetric gravity, until a few years ago a way to avoid the ghost was discovered, by composing the interaction of the two metrics out of a very specific set of symmetric potential terms. These potentials are most elegantly understood in the tetrad formalism as the five possible invariant volume elements one can construct from the two sets of tetrads available~\cite{deRham_et_Gabadadze2010b,deRham_Gabadadze_et_Tolley2010,deRham_Gabadadze_et_Tolley2011b,deRham_Gabadadze_et_Tolley2011c,
Hassan_et_Rosen2011a,Hassan:2011hr} (see also Refs. \cite{Golovnev:2011aa,Hassan:2011ea,Hassan:2012qv,Kluson:2012wf,Nomura:2012xr,Golovnev:2013fj,Kluson:2013cy,Golovnev:2014upa}, and Ref. \cite{deRham:2014zqa} for a comprehensive and readable recent review on massive gravity and its bimetric generalizations). There are indications that these terms are the only ones that are allowed in order for the theory to be ghost-free; in particular, kinetic interactions between the two metrics do not appear to be consistent \cite{deRham:2013tfa,Gao:2014jja}. In the most promising version of the resultant theories, both metrics (or the tetrads associated to them) are dynamical~\cite{Hassan_et_al2012a,Hassan_et_Rosen2012a,Hassan:2011ea}, leaving us with a theory of massive bigravity.

Thus we have at hand a unique class of viable bimetric theories,\footnote{Possible problems with, e.g., superluminal propagation may need further study \cite{Deser:2013qza,Deser:2013rxa}.} completely specified up to the magnitudes of the three interaction terms and two cosmological constants which cannot presently be inferred from first principles.  An ambiguity arises in the presence of matter fields. They could in principle couple to both of the metrics, with the relative coupling strength introducing an additional free parameter. Here, however, we consider the ``singly-coupled'' realization of the theory in which matter is only coupled to one of the metrics. This metric then retains its usual role as the physical spacetime metric, used for measuring distances and time intervals, while the other metric can be interpreted as an additional tensor field whose nonminimal interactions with the first metric gives the graviton a nonzero mass. (A massless spin-2 mode also propagates; matter couples to a combination of the two modes.)

Matter could also in principle be allowed to couple to both metrics \cite{Akrami:2013ffa,Tamanini:2013xia,Akrami:2014lja}. Such a multimetric matter coupling, also mentioned in Refs. \cite{Khosravi:2011zi,Hassan_et_Rosen2012a}, extends the metric-exchange symmetry of the vacuum theory to the full theory, and has interesting implications for the effective nature of spacetime \cite{Akrami:2014lja}. Alternatively, one may double the matter content and consider two separate sectors of matter fields, each coupled exclusively to one of the metrics \cite{Baccetti:2012bk,Baccetti:2012re,Capozziello:2012re,Comelli:2014bqa,DeFelice:2014nja}. Note, however, that many such doubly-coupled theories reintroduce the Boulware-Deser ghost \cite{Yamashita:2014fga,deRham:2014naa}, and the identification of a ghost-free double coupling requires further study. A theory has recently been constructed where matter couples to an effective metric built out of both metrics \cite{deRham:2014naa,Noller:2014sta,Schmidt-May:2014xla}. This new double coupling has interesting cosmological implications \cite{Enander:2014xga,Gumrukcuoglu:2014xba,mgletter}, is ghost-free up to at least the strong-coupling scale \cite{deRham:2014naa,deRham:2014fha}, and is potentially ghost-free at all energies \cite{Hassan:2014gta}. Several other generalizations of the theory in which new fields are introduced have also been proposed in, for example, Refs. \cite{Nojiri:2012zu,Nojiri:2012re,Paulos_et_Tolley2012,Cai:2012ag,Huang:2012pe,Wu:2013ii,Leon:2013qh}. Various other recent considerations on bimetric spaces include Refs. \cite{Hossenfelder:2008bg,Amendola:2010bk,BeltranJimenez:2012sz}, and biconnected spacetimes have been studied in Refs. \cite{Koivisto:2011vq,Tamanini:2012mi,Khosravi:2013kha}. Finally, let us note that the graviton could acquire a mass without the presence of an additional metric if locality is violated; see Refs. \cite{Jaccard:2013gla,Maggiore:2013mea,Nesseris:2014mea} and 
\cite{Maggiore:2014sia,Modesto:2013jea} for studies of two interesting realizations.\footnote{It remains to be understood whether these theories can be directly related to bimetric gravity as an effective theory after integrating out one of the metrics, along the lines of Ref. \cite{Hassan:2013pca}. Initial steps in this direction --- up to quadratic order in curvature invariants and taking the subset of bimetric gravity admitting flat-space vacua --- have been taken in Ref. \cite{Cusin:2014zoa}.}

In this article we undertake a study of the cosmological large-scale structures (LSS) in the singly-coupled version of massive bimetric gravity, with the aim of understanding the deviations from the predictions of general relativity (GR) and the potentially testable cosmological signatures of the bimetric nature of gravity. We are motivated in particular by anticipation of the forthcoming Euclid mission which is expected to improve the accuracy of the present large-scale structure data by nearly an order of magnitude \cite{Laureijs:2011gra,Amendola:2012ys}. The cosmological background expansion in the bimetric theory has been solved and found to naturally produce the observed late-time acceleration of the Universe without introducing an explicit cosmological constant~\cite{Volkov:2011an,Comelli:2011zm,vonStrauss:2011mq,Volkov2012a,Volkov:2012zb,Berg:2012kn,Akrami:2012vf,Akrami:2013pna}. Thus the dark energy problem could be connected to a finite mass of the graviton. The theory has also been shown to be consistent with the strong lensing properties of elliptical galaxies \cite{Enander:2013kza}.

Parameter constraints have been derived by careful statistical analysis comparing the model predictions with the available background cosmological data, and the bimetric models have been shown to provide as good a fit as the standard $\Lambda$CDM model, including one case (the minimal $\beta_1$-only model) which has the same number of free parameters as $\Lambda$CDM \cite{Akrami:2012vf,Akrami:2013pna,Konnig:2013gxa}. However, the parameter constraints from the background expansion data have strong degeneracies within the theory itself and furthermore, as is well-known, a given background expansion can be reproduced by many different theories. To efficiently test the theory and distinguish its cosmology from others one needs to study the formation of structures. The first steps in this task were taken by Berg {\it et al.} \cite{Berg:2012kn}, who derived the linear cosmological perturbation equations. Here we derive and study the (equivalent) set of equations in general terms and, more importantly, proceed to make the link to observable quantities. Recently, K\"onnig and Amendola explored large-scale structures in a one-parameter bigravity model \cite{Konnig:2014dna}. Our results are in complete agreement with theirs. Cosmological perturbations in bimetric gravity have also been studied in Refs. \cite{Comelli:2012db,Comelli:2014bqa,DeFelice:2014nja}. Note that these references claim the existence of an instability in the evolution of perturbations. Since the initial submission of this work, a subset of bigravity models with early-time instabilities were uncovered in Ref. \cite{Konnig:2014xva}. This would invalidate the use of linear perturbation theory in those specific models at those early times. We will present our results for all models, stable or not, for illustrative purposes, and mention the stability properties when relevant.

This article is organized as follows. In \cref{bigravity} we review the bimetric massive gravity theory and its cosmological self-accelerating background solutions. We then present the cosmological perturbation equations in \cref{subhorizon} and focus on the formation of structures in the matter-dominated era. The observationally relevant part of the matter power spectrum is at the subhorizon scales, and in this limit we arrive at a convenient closed-form evolution equation in \cref{results} that captures the modifications to the general relativistic growth rate of linear structures, as well as the leading order scale-dependence which modifies the shape of the spectrum at near-horizon scales. The full perturbation equations are detailed in \cref{app:perteqs}, and the general form of the coefficients in the closed-form equation is given in \cref{app:hcoeff}. The results are analyzed numerically in \cref{results}, where we discuss the general features of the models and confront them with the observational data. We conclude in \cref c.

\section{Theory and cosmological framework}
\label{bigravity}

\subsection{Massive bigravity}

The ghost-free theory of massive bigravity is defined by the action \cite{Hassan_et_Rosen2012a}
\begin{align}
  S &= - \frac{M^2_g}{2}\int d^4x\sqrt{-\det g}R(g) - \frac{M^2_f}{2}\int d^4x\sqrt{-\det f}R(f) \nonumber\\
  &\hphantom{{}=} + m^2M_g^2\int d^4x\sqrt{-\det g}\sum_{n=0}^{4}\beta_ne_n\left(\sqrt{g^{-1}f}\right) \nonumber\\
  &\hphantom{{}=} + \int d^4x\sqrt{-\det g}\mathcal{L}_m\left(g, \Phi\right), \label{eq:action} 
\end{align}
where $g_{\mu\nu}$ and $f_{\mu\nu}$ are spin-2 tensor fields with metric properties. They are dynamical, and they define the gravitational sector of the theory. For conciseness we will frequently refer to the metrics as $g$ and $f$, suppressing their indices. Their kinetic terms are the usual Einstein-Hilbert terms, and the fields interact through a potential comprising five terms which are particular functions of the fields $g$ and $f$ (but not their derivatives):
\begin{align}
e_0\left(\mathbbm{X}\right) &\equiv 1, \qquad \nonumber \\
e_1\left(\mathbbm{X}\right) &\equiv \left[\mathbbm{X}\right], \qquad \nonumber \\
e_2\left(\mathbbm{X}\right) &\equiv \frac{1}{2}\left(\left[\mathbbm{X}\right]^2 - \left[\mathbbm{X}^2\right]\right),\nonumber \\
e_3\left(\mathbbm{X}\right) &\equiv \frac{1}{6}\left(\left[\mathbbm{X}\right]^ 3 - 3\left[\mathbbm{X}\right]\left[\mathbbm{X}^2\right] + 2\left[\mathbbm{X}^3\right]\right), \qquad \nonumber \\
e_4\left(\mathbbm{X}\right) &\equiv \det\left(\mathbbm{X}\right).
\end{align}
Here, $e_n(\mathbbm{X})$ are the elementary symmetric polynomials of the eigenvalues of the matrix $\mathbbm{X}\equiv\sqrt{g^{-1}f}$, and the square brackets $[\mathbbm{X}]$, $[\mathbbm{X}^2]$, and $[\mathbbm{X}^3]$ denote traces of the matrices $\mathbbm{X}$, $\mathbbm{X}^2$, and $\mathbbm{X}^3$, respectively. The quantities $\beta_n$ $(n=0,...,4)$ are the free parameters of the theory. Notice that $\beta_0$ and $\beta_4$ are just cosmological constant terms for the $g$ and $f$ metrics, respectively. $M_g$ and $M_f$ are the two Planck masses for $g$ and $f$, respectively. The parameter $m$ is not an independent free parameter, although we will keep it explicit because it will help to keep track of the interaction terms in all equations. Finally, the last line of the action (\ref{eq:action}) contains the usual matter sector, minimally coupled to $g_{\mu\nu}$. The action (\ref{eq:action}) has the useful property that, excluding the matter coupling, it is symmetric \cite{Hassan_et_Rosen2012a,Hassan:2012wr} under the exchange of the two metrics up to parameter redefinitions,
\begin{equation}
g_{\mu\nu} \leftrightarrow f_{\mu\nu},\qquad \beta_n \rightarrow \beta_{4-n},\qquad M_g \leftrightarrow M_f. \label{eq:symmetry}
\end{equation}

The two interacting spin-2 fields have seven propagating degrees of freedom, corresponding to a massless and a massive graviton (two and five degrees of freedom, respectively) \cite{Hassan:2012wr}. It has been shown to be free of the Boulware-Deser ghost instability \cite{Hassan_et_Rosen2012a}. In the gravitational sector, the action (\ref{eq:action}) represents the most general ghost-free theory of two interacting spin-2 fields with interaction (mass) terms that are functions of the fields only (i.e., not of their derivatives).\footnote{Indeed, it has been suggested that this is the most general ghost-free theory even when derivative interactions are allowed \cite{deRham:2013tfa}.}

The existence of a consistent bimetric theory raises an intriguing question: which is the physical metric? In the original theory of ghost-free massive bigravity, only one of the two spin-2 fields, which we take to be $g$, directly couples to matter, while the other dynamical spin-2 field, $f$, only interacts with matter fields indirectly through its interactions with $g$.\footnote{From this perspective, massive bigravity is not dissimilar to other commonly-considered modifications of GR, such as scalar-tensor theories \cite{Horndeski:1974wa,Deffayet:2011gz} or vector-tensor theories \cite{Jacobson:2008aj,Solomon:2013iza}, which have a single physical metric coupled nonminimally to some other field or fields.} It is therefore natural to interpret $g$ in this case as the usual ``metric" of spacetime, while $f$ is an extra spin-2 field that is required in order to give mass to the graviton. This is sometimes referred to as the ``singly-coupled" theory of massive bigravity. Crucially, it is only in this case that the ghost-free nature of bigravity has been proven, and indeed many extensions to couple matter to both metrics bring back the Boulware-Deser ghost \cite{Yamashita:2014fga,deRham:2014naa}. Both for simplicity and to be certain we are working with a (Boulware-Deser) ghost-free theory, we restrict ourselves in this paper to the singly-coupled theory. However, since the fields $g$ and $f$ have metric properties, we will call both $g$ and $f$ ``metrics," even though strictly speaking, the singly-coupled theory could more accurately be called a theory of ``gravity coupled to matter and a symmetric 2-tensor"~\cite{Berg:2012kn}.

By varying the action (\ref{eq:action}) with respect to $g$ and $f$ we obtain the generalized Einstein equations for the two metrics,
\begin{align}
R_{\mu\nu}^g - \frac{1}{2}g_{\mu\nu} R^g + m^2\sum_{n=0}^3\left(-1\right)^{n}\beta_ng_{\mu\lambda}Y^{\lambda}_{(n)\nu}\left(\sqrt{g^{-1}f}\right) &= \frac{1}{M_g^2}T_{\mu\nu}, \label{eq:Einsteingeng}\\
R_{\mu\nu}^f - \frac{1}{2}f_{\mu\nu} R^f + \frac{m^2}{M^2_\star}\sum_{n=0}^3\left(-1\right)^{n}\beta_{4 - n}f_{\mu\lambda}Y^{\lambda}_{(n)\nu}\left(\sqrt{f^{-1} g}\right) &= 0, \label{eq:Einsteingenf}
\end{align}
where $R_{\mu\nu}^g$ and $R_{\mu\nu}^f$ are the Ricci tensors and $R^g$ and $R^f$ are the Ricci scalars corresponding to the metrics $g$ and $f$, respectively. We have defined $M_\star^2 \equiv M_f^2/M_g^2$. The functions $Y_{(n)}(\mathbbm{X})$ are defined as
\begin{align}
  Y_{(0)}\left(\mathbbm{X}\right) &\equiv \mathbbm{1}, \quad \nonumber \\
  Y_{(1)}\left(\mathbbm{X}\right) &\equiv \mathbbm{X} - \mathbbm{1}\left[\mathbbm{X}\right],\quad \nonumber \\
  Y_{(2)}\left(\mathbbm{X}\right) &\equiv \mathbbm{X}^2 - \mathbbm{X}\left[\mathbbm{X}\right] + \frac{1}{2}\mathbbm{1}\left(\left[\mathbbm{X}\right]^2 - \left[\mathbbm{X}^2\right]\right), \nonumber \\  
  Y_{(3)}\left(\mathbbm{X}\right) &\equiv \mathbbm{X}^3 - \mathbbm{X}^2\left[\mathbbm{X}\right] + \frac{1}{2}\mathbbm{X}\left(\left[\mathbbm{X}\right]^2 - \left[\mathbbm{X}^2\right]\right) \nonumber \\
  &\hphantom{{}\equiv} - \frac{1}{6}\mathbbm{1}\left(\left[\mathbbm{X}\right]^ 3 - 3\left[\mathbbm{X}\right]\left[\mathbbm{X}^2\right] + 2\left[\mathbbm{X}^3\right]\right). \label{eq:ymatdef}
\end{align}
The tensors $g_{\mu\lambda}Y^{\lambda}_{(n)\nu}(\sqrt{g^{-1}f})$ and $f_{\mu\lambda}Y^{\lambda}_{(n)\nu}(\sqrt{f^{-1} g})$ are symmetric \cite{Hassan:2012wr}. Finally, $T_{\mu\nu}$ is the stress-energy tensor defined with respect to the physical metric $g$,
\begin{equation}
T_{\mu\nu} \equiv - \frac{2}{\sqrt{-\det g}}\frac{\delta(\sqrt{-\det g}\mathcal{L}_m^g)}{\delta g^{\mu\nu}}.
\end{equation}

General covariance of the matter sector implies conservation of the stress-energy tensor, as in GR:
\begin{equation}\label{eq:ParticleEOM}
\nabla_g^\mu T_{\mu\nu} = 0.
\end{equation}
Furthermore, by combining the Bianchi identities for the $g$ and $f$ metrics with the field equations (\ref{eq:Einsteingeng}) and (\ref{eq:Einsteingenf}), we obtain the following two Bianchi constraints on the mass terms:
\begin{align}
 \nabla_g^{\mu}\frac{m^2}{2}\sum_{n=0}^3\left(-1\right)^{n}\beta_ng_{\mu\lambda}Y^{\lambda}_{(n)\nu}\left(\sqrt{g^{-1}f}\right) &= 0, \label{eq:Bianchig}\\
 \nabla_f^{\mu}\frac{m^2}{2M^2_\star}\sum_{n=0}^3\left(-1\right)^{n}\beta_{4 - n}f_{\mu\lambda}Y^{\lambda}_{(n)\nu}\left(\sqrt{f^{-1} g}\right) &= 0, \label{eq:Bianchif}
\end{align}
after using \cref{eq:ParticleEOM}.

Finally we note that, as discussed in \cite{Berg:2012kn,Akrami:2012vf}, we can constantly rescale the $f$ metric and $\beta_i$ parameters in such a way that only $\Ms$ changes. This means that $\Ms$ is redundant and is not a free parameter. Herein we use this freedom to set $\Ms=1$.

\subsection{Background cosmology and self-acceleration}
\label{background}

In this subsection we review the homogeneous and isotropic cosmology of massive bigravity. Following Refs. \cite{vonStrauss:2011mq,Akrami:2012vf,Akrami:2013ffa} we assume that, at the background level, the Universe can be described by Friedmann-Lema\^{\i}tre-Robertson-Walker (FLRW) metrics for both $g$ and $f$. Specializing to a spatially flat universe and working in conformal time, we have
\begin{align}
ds_g^2 &= a^2 \left(-d\tau^2 + d\vec{x}^2\right), \label{eq:frwg}\\
ds_f^2 &= -X^2 d\tau^2 + Y^2 d\vec{x}^2. \label{eq:frwf}
\end{align}
Here, $a$ and $Y$ are the spatial scale factors for $g$ and $f$, respectively, and are functions of time. $X$ is the lapse function for the metric $f$ and is also a function of time.

With these choices of metrics, the generalized Einstein equations (\ref{eq:Einsteingeng}) and (\ref{eq:Einsteingenf}), assuming a perfect-fluid source with density $\rho = -{T^0}_0$, give rise to two Friedmann equations \cite{Akrami:2012vf},
\begin{align}
3H^2 - m^2a^2\left(\beta_0 + 3\beta_1y + 3\beta_2y^2 + \beta_3y^3\right) &= \frac{a^2\rho}{M_g^2}, \label{eq:friedmann_g}\\
3K^2 - m^2X^2\left(\beta_1y^{-3} + 3\beta_2y^{-2} + 3\beta_3y^{-1} + \beta_4\right) &= 0,\label{eq:friedmann_f}
\end{align}
where $y \equiv Y/a$, $H \equiv \dot{a}/a$, $K \equiv \dot{Y}/Y$, and $\dot{} = d/d\tau$. Note that $H$ is the conformal-time, not cosmic-time, Hubble parameter. While all the expressions we present in this section hold for any perfect fluid, we will specialize in the following sections to pressureless dust, which obeys
\begin{equation}
\dot\rho + 3H\rho = 0. \label{eq:rhocons}
\end{equation}
The Bianchi constraints imply that $X$ can be written in terms of $Y$, $H$, and $K$ as\footnote{The Bianchi identites are also satisfied if $\beta_1 + 2\beta_2y + \beta_3y^2 = 0$. This is not generic as it requires tuned initial conditions. Moreover, it leads to cosmologies which are exactly $\Lambda$CDM at the background level \cite{Comelli:2011zm}, which we would consider less interesting than the cases studied in this work. Hence we will assume that this condition is not satisfied, and \cref{eq:xkyh} holds.}
\begin{align}
X &= \frac{KY}{H}.\label{eq:xkyh}
\end{align}

The Friedmann equations, (\ref{eq:friedmann_g}) and (\ref{eq:friedmann_f}), and the Bianchi identity (\ref{eq:xkyh}) can be combined to find an algebraic, quartic equation for $y$,
\begin{equation}
\beta_3y^4 + \left(3\beta_2 - \beta_4\right)y^3 + 3\left(\beta_1 - \beta_3\right)y^2 + \left(\frac{\rho}{M_g^2m^2} + \beta_0 - 3\beta_2\right)y - \beta_1 = 0.\label{eq:quartic}
\end{equation}
Using this quartic equation and the $g$-metric Friedmann equation (\ref{eq:friedmann_g}), as well as the fluid conservation equation (\ref{eq:rhocons}), we can express all background quantities solely in terms of $a(\tau)$ and $y(\tau)$:
\begin{align}
\frac{\dot y}{y} &= -3H\frac{\beta_3y^4 + (3\beta_2 - \beta_4)y^3 + 3(\beta_1 - \beta_3)y^2 + (\beta_0 - 3\beta_2)y - \beta_1}{3\beta_3y^4 + 2(3\beta_2-\beta_4)y^3 + 3(\beta_1-\beta_3)y^2 + \beta_1}, \label{eq:brepyp} \\
H^2 &= m^2a^2\left(\frac{\beta_4}{3}y^2 + \beta_3 y + \beta_2 + \frac{\beta_1}{3}y^{-1}\right), \label{eq:brepH} \\
K &= H + \frac{\dot y}{y}, \label{eq:brepK} \\
\frac{\rho}{m^2M_g^2} &= -\beta_3y^3 + (\beta_4 - 3\beta_2)y^2 + 3(\beta_3 - \beta_1)y + 3\beta_2 - \beta_0 + \beta_1y^{-1}. \label{eq:breprho}
\end{align}
These will be crucial in the rest of this paper, since they reduce the problem of finding any parameter --- background or perturbation --- to solving \cref{eq:brepyp} for $y(z)$, where $z=1/a-1$ is the redshift.

As discussed in detail in Ref. \cite{Akrami:2012vf}, \cref{eq:friedmann_g,eq:quartic} completely determine the expansion of the Universe. In the same study, the expansion history was compared to background observations, placing constraints on the free parameters, $m^2\beta_i$. This analysis showed that massive bigravity is able to account for the late-time acceleration of the Universe in agreement with present observations without resorting to a cosmological constant or vacuum energy, i.e., without turning on $\beta_0$.\footnote{Some treatments of massive and bimetric gravity use an alternative set of parameters, $\alpha_i$, which are related linearly to the $\beta_i$ parameters \cite{Hassan_et_Rosen2011a}. It is sometimes claimed (see, e.g., section 6 of Ref. \cite{deRham:2014zqa}) that $\alpha_0$ is the $g$-metric cosmological constant. To a certain extent this is a semantic distinction, but we choose to consider $\beta_0$ the cosmological constant, since it is this parameter which receives contributions from matter loops. See footnotes 5 and 8 of Ref. \cite{Akrami:2012vf} for an involved discussion on this point.} In other words, the theory provides self-accelerating solutions which yield a dynamical dark energy (i.e., a varying cosmological constant)\footnote{The singly-coupled version of the theory cannot mimic an exact $\Lambda$CDM universe for generic initial conditions, although by additionally coupling matter minimally to the $f$ metric there are regions of parameter space which are exactly $\Lambda$CDM at the level of the background \cite{Akrami:2013ffa}.} in excellent agreement with observations without an explicit cosmological constant. One of the reasons to go beyond background observations and towards cosmic structure is to break this degeneracy between bigravity and $\Lambda$CDM.

One simple, but interesting, sub-model of the full bigravity theory is the case where all of the $\beta$ parameters (including the cosmological terms $\beta_0$ and $\beta_4$) are set to zero except for $\beta_1$. (We emphasize here, however, that $\beta_4$ is not a cosmological term in the usual sense, as a $\beta_4$-only model will not have any physical effects, and this term does not receive contributions from the vacuum energy.) It has been shown in Refs. \cite{Akrami:2012vf,Konnig:2013gxa} that this model provides a self-accelerating evolution which agrees with background cosmological observations and, as it possesses the same number of free parameters as the standard $\Lambda$CDM model, is a viable alternative to it.\footnote{Indeed, it may be more viable than $\Lambda$CDM if the graviton mass turns out to be stable to quantum corrections, as is suggested by, e.g., Ref. \cite{deRham:2013qqa}. Famously, a small cosmological constant is highly unstable to quantum effects.} The graviton mass in this case is given by $\sqrt{\beta_1}m$. Note that in order to give rise to acceleration at the present era, the graviton mass typically should be comparable to the present-day Hubble rate, $\beta_1m^2\sim H_0^2$.

The minimal $\beta_1$-only model is also distinctive for having a phantom equation of state, $w(z)\approx-1.22_{-0.02}^{+0.02}-0.64_{-0.04}^{+0.05}z/(1+z)$ at small redshifts;\footnote{This prediction for $w(z)$ is based on a best-fit value $m^2\beta_1/H_0^2 = 1.38 \pm 0.03$ \cite{Konnig:2013gxa}. Our best fit, $m^2\beta_1/H_0^2=1.45 \pm 0.02$ (see Ref. \cite{Akrami:2012vf} and \cref{sec:b1} of this work), would imply a slightly different value for $w(z)$.} moreover, $w$ is related in a simple way to the matter density parameter \cite{Konnig:2013gxa}. This provides a concrete and testable prediction of the model that can be verified by future LSS experiments, such as Euclid \cite{Laureijs:2011gra,Amendola:2012ys}, intensity mappings of neutral hydrogen \cite{Chang:2007xk,Bull:2014rha}, and combinations of LSS and cosmic microwave background (CMB) measurements \cite{Majerotto:2012mf}. The model has also been proven in \cite{Fasiello:2013woa} to satisfy an important stability bound, and is therefore a compelling candidate for explaining the cosmic acceleration. It is, however, worth noting that its linear cosmological perturbations are unstable until $z\sim0.5$, as shown in Ref. \cite{Konnig:2014xva}: as emphasized in that work, this fact does not rule out this model, but does impede our ability to use linear perturbation theory to describe perturbations at all times. This raises the interesting question of how to make predictions for structure formation during the unstable period.

The aforementioned studies have largely been restricted to the background cosmology of the theory. As the natural next step, in the rest of the present work we will extend the predictions of massive bigravity to the perturbative level, and study how consistent the models are with the observed growth of structures in the Universe. We restrict ourselves to the linear, subhorizon r\'egime and examine whether there are any deviations from the standard model predictions which may be observable by future LSS experiments. Note that due to the aforementioned instability, some (though not all) bimetric models cannot be treated perturbatively; as such, structure formation must be studied nonlinearly in these models. We choose to study every model with a sensible background evolution and one or two free $\beta_i$ parameters, and so some unstable cases will be included. These should be seen as toy models, useful for illustrative purposes; throughout the paper we point out which theories do and do not suffer from this instability.

\section{Linear perturbations, subhorizon limit and the choice of gauge}
\label{subhorizon}

We define the perturbations to the FLRW metrics by extending \cref{eq:frwg,eq:frwf} to
\begin{align}
ds_g^2 &= a^2\left\{-(1+E_g)d\tau^2 + 2\partial_iF_gd\tau dx^i + \left[(1+A_g)\delta_{ij} + \partial_i\partial_jB_g\right]dx^idx^j\right\}, \\
ds_f^2 &= -X^2(1+E_f)d\tau^2 + 2XY\partial_iF_fd\tau dx^i + Y^2\left[(1+A_f)\delta_{ij} + \partial_i\partial_jB_f\right]dx^idx^j,
\end{align}
where the perturbations $\{E_{g,f},A_{g,f},F_{g,f},B_{g,f}\}$ are allowed to depend on both time and space. Spatial indices are raised and lowered with the Kronecker delta. The stress-energy tensor is defined up to linear order by
\begin{align}
T{}^0{}_0 &= -\bar\rho(1+\delta), \nonumber \\
T{}^i{}_0 &= -\left(\bar\rho + \bar P\right)v^i, \nonumber \\
T{}^0{}_i &= \left(\bar\rho + \bar P\right)\left(v_i + \partial_iF_g\right), \nonumber \\
T{}^i{}_j &= \left(\bar P + \delta P\right)\delta^i{}_j + \Sigma{}^i{}_j, \label{eq:pertstresstens}
\end{align}
where $v^i \equiv dx^i/dt$ and $\Sigma{}^i{}_i=0$. We specialize immediately to dust ($P=\delta P = \Sigma^i{}_j = 0$) and define the velocity divergence $\theta \equiv \partial_i v^i$.

We derive the perturbed Einstein and conservation equations in \cref{app:perteqs}. These equations are quite complicated; in order to isolate the physics of interest, namely that of linear structures in the subhorizon r\'egime, we focus on the subhorizon (or quasistatic) limit of the field equations. This limit is defined by taking $k^2\Phi\gg H^2\Phi\sim H\dot\Phi \sim \ddot\Phi$ for any variable $\Phi$, where we have expanded in Fourier modes with wavenumber $k$, and assuming $K \sim H$ so we can take the same limit in the $f$-metric equations. Moreover, we will take $A_{g,f}$ and $E_{g,f}$ to be of the same order as $kF_{g,f}$ and $k^2B_{g,f}$, as these terms all appear this way in the linearized metric. Finally, we will work in Newtonian gauge for the $g$ metric, defined by setting $F_g = B_g = 0$. In the singly-coupled theory where $g$ is the ``physical" metric, i.e., only $g$ couples (minimally) to matter, this is as sensible a gauge choice as it is in GR.\footnote{Given two separate diffeomorphisms for the $g$ and $f$ metrics, only the diagonal subgroup of the two preserves the mass term. In practice, this means that we have a single coordinate system which we may transform by infinitesimal diffeomorphisms, exactly as in GR.}

With these definitions, we can write down the subhorizon evolution equations. The energy constraint (the $0-0$ Einstein equation) for the $g$ metric is
\begin{equation}
\left(\frac{k}{a}\right)^2 \left(A_g + \frac{m^2}{2}yPa^2B_f\right) + \frac{3}{2}m^2yP\left(A_g-A_f\right) = \frac{\bar\rho}{M_g^2}\delta. \label{eq:00g}
\end{equation}
The trace of the $i-j$ equation yields
\begin{equation}
\left(\dot H - H^2 + \frac{a^2\bar\rho}{2M_g^2}\right)E_g + m^2a^2\left[\frac{1}{2}xP\left(E_f - E_g\right) + yQ\left(A_f - A_g\right)\right] = 0. \label{eq:iig}
\end{equation}
The off-diagonal piece of the $i-j$ equation tells us
\begin{equation}
A_g + E_g + m^2a^2yQB_f = 0. \label{eq:ijg}
\end{equation}
There is, additionally, the momentum constraint (the $0-i$ Einstein equation), which we have used along with the velocity conservation equation (\ref{eq:thetacons}) to simplify the trace $i-j$ equation (\ref{eq:iig}).\footnote{In the approach of Ref. \cite{Konnig:2014dna}, where a slightly different limit is taken by dropping all derivative terms, this step is crucial for the results to be consistent; in our case it is simply useful for rewriting derivatives of $\dot A_g$ and $\ddot A_g$ in terms of $E_g$, so that the equation is manifestly algebraic in the perturbations.} Note also that we have used the off-diagonal piece, \cref{eq:ijg}, to eliminate some redundant terms in the trace equation.
In the above, we have made the definitions
\begin{align}
P &\equiv \beta_1 + 2\beta_2y + \beta_3y^2, \label{eq:Pdef} \\
Q &\equiv \beta_1 + \left(x + y\right)\beta_2 + xy\beta_3, \label{eq:pertQdef} \\
x &\equiv X/a, \\
y &\equiv Y/a.
\end{align}

For the $f$ metric, the corresponding equations are
\begin{align}
\left(\frac{k}{a}\right)^2 \left(A_f - \frac{m^2}{2}\frac{Pa^2}{y}B_f\right) + \frac{3m^2}{2}\frac{P}{y}\left(A_f - A_g\right) &= 0, \label{eq:00f} \\
\left[-\dot K + \left(H + \frac{\dot x}{x}\right)K\right]E_f + m^2\frac{a^2x}{y^2}\left[\frac{1}{2}P\left(E_f - E_g\right) + Q\left(A_f - A_g\right)\right] &= 0,
 \label{eq:iif} \\
A_f + E_f - m^2\frac{Qa^2}{x}B_f &= 0. \label{eq:ijf}
\end{align}
The fluid conservation equations are unchanged from GR,
\begin{align}
\dot\delta + \theta &=0, \label{eq:deltacons} \\
\dot\theta + H\theta - \frac{1}{2}k^2 E_g &= 0. \label{eq:thetacons}
\end{align}

In GR the trace equation, (\ref{eq:iig}), adds no new information: it becomes $0=0$ after using the Friedmann equation. In massive bigravity this equation does carry information and it is crucial that we use it. However, we can still simplify it by using the background equations, obtaining\footnote{This equation holds beyond the subhorizon limit, in a particular gauge; see \cref{app:perteqs}.}
\begin{equation}
m^2\left[P\left(xE_f-yE_g\right)+2yQ\Delta A\right]=0. \label{eq:iig-simplified}
\end{equation}
Note that the $m\to0$ limit gives $0=0$, as expected. There is one interesting feature: if we substitute the background equations into the $f$-metric trace equation, (\ref{eq:iif}), we obtain \cref{eq:iig-simplified} again. Hence one of the two trace equations is redundant and can be discarded; so what looks like a system of six equations is actually a system of five.

With these relations, the system of equations presented in this section is closed.

\section{Results: growth of structures and cosmological observables}
\label{results}

In this section we study the linearized growth of subhorizon structures, first solving the Einstein equations for these structures and then comparing to data. Deviations from these relations due to modified gravity can be summarized by a few parameters which are observable by large-scale structure surveys such as Euclid \cite{Laureijs:2011gra,Amendola:2012ys}. The main aim of this section is to see under what circumstances these parameters are modified in the linear r\'{e}gime by massive bigravity.

\subsection{Modified gravity parameters}

We will focus on three modified growth parameters, defined in the Euclid Theory Working Group review \cite{Amendola:2012ys} as $f$ (and its parametrization $\gamma$), $Q$, and $\eta$. They are:
\begin{description}
\item[Growth rate ($f$) and index ($\gamma$):] These parameters measure the growth of structures, and are defined by
\begin{equation}
f(a,k) \equiv \frac{d\log \delta}{d\log a} \approx \Omega_m^\gamma, \label{eq:gammadef}
\end{equation}
where $\Omega_m \equiv a^2\bar\rho/(3M_g^2H^2)$ is the usual matter density parameter.
\item[Modification of Newton's Constant ($Q$):] The function $Q(a,k)$\footnote{Not to be confused with the $Q$ defined in \cref{eq:pertQdef}.} parametrizes modifications to Newton's constant in the Poisson equation,
\begin{equation}
\frac{k^2}{a^2}A_g \equiv \frac{Q(a,k)\bar\rho}{M_g^2}\delta. \label{eq:mgQdef}
\end{equation}
\item[Anisotropic Stress ($\eta$):] Effective anisotropic stress leads the quantity $A_g + E_g$ to deviate from its GR value of zero, which we can parametrize by the parameter $\eta(a,k)$:
\begin{equation}
\eta(a,k) \equiv -\frac{A_g}{E_g}. \label{eq:etadef}
\end{equation}
\end{description}
In GR, these parameters have the values $\gamma \approx 0.545$ and $Q = \eta = 1$.

We have five independent Einstein equations (\cref{eq:00g,eq:ijg,eq:00f,eq:ijf,eq:iig-simplified}) for five metric perturbations\footnote{After gauge fixing there are six metric perturbations, but once we substitute the $0-i$ equations into the trace $i-j$ equations, $F_f$ drops out of our system. In a gauge where $F_g=F_f=0$, as was used in Ref. \cite{Konnig:2014dna}, the equivalent statement is that the $B_g$ and $B_f$ parameters are only determined up to their difference, $B_f - B_g$, which is gauge invariant.} and $\delta$. Crucially, this system is algebraic. There are five equations for six variables, so we can only solve for any five of the perturbations in terms of the sixth. Of the modified growth parameters, $Q$ and $\eta$ are ratios of perturbations so are insensitive to how we solve the system. However, to find $\gamma$ we need to solve a differential equation for $\delta$. It is therefore simplest to solve for the perturbations $\{A_{g,f},E_{g,f},B_f\}$ in terms of $\delta$.

Solving the system, we find each perturbation can be written in the form $f(\tau,k)\delta$, for some function $f(t)$. We do not display the solutions here as they are quite unwieldy, although we do note that in the limit with only $\beta_1 \neq 0$ studied in Ref. \cite{Konnig:2014dna}, and taking into account differences in notation and gauge, our expressions for the perturbations match theirs.\footnote{We thank the authors of Ref. \cite{Konnig:2014dna} for providing us with these expressions, not all of which appeared in their paper.}

With these solutions for $\{A_{g,f},E_{g,f},B_f\}$ in hand, we can immediately read off $Q$ and $\eta$. To calculate the growth index, $\gamma$, we need to solve a conservation equation for the density contrast, $\delta$. The fluid conservation equations, (\ref{eq:deltacons}) and (\ref{eq:thetacons}), are unchanged from GR, so as in GR we can manipulate them to find the usual evolution equation for $\delta$ sourced by the gravitational potential,
\begin{equation}
\ddot \delta + H\dot\delta + \frac{1}{2}k^2E_g(\delta) = 0. \label{eq:deltaevol}
\end{equation}
At this point we diverge from the usual story. In GR, there is no anisotropic stress and the Poisson equation holds, so $k^2E_g = -(a^2\bar\rho/M_g^2)\delta$. Both of these facts are changed in massive bigravity, so there is a modified (and rather more complicated) relation between $k^2E_g$ and $\delta$. However, since we do have such a relation, $\delta$ still obeys a closed second-order equation which we can solve numerically.

Finally, we note that the three modified gravity parameters are encapsulated by five time-dependent parameters. The expressions for $Q$ and $\eta$ can be written in the form
\begin{align}
\eta &= h_2\left(\frac{1 + k^2h_4}{1+k^2h_5}\right), \label{eq:expreta} \\
Q &= h_1\left(\frac{1 + k^2h_4}{1+k^2h_3}\right), \label{eq:exprQ}
\end{align}
where the $h_i$ are functions of time only and depend on $m^2\beta_i$. The same result has been obtained for Horndeski gravity \cite{DeFelice:2011hq,Amendola:2012ky}, which is the most general scalar-tensor theory with second-order equations of motion \cite{Horndeski:1974wa}. This similarity was noticed in the $\beta_1$-only model in Ref. \cite{Konnig:2014dna}, where it was attributed to the fact that the field equations of ghost-free massive bigravity are also second order.

Furthermore, the structure growth equation, (\ref{eq:deltaevol}), can be written in terms of $Q$ and $\eta$ and hence the $h_i$ coefficients as
\begin{equation}
\ddot \delta + H\dot\delta - \frac{1}{2}\frac{Q}{\eta}\frac{a^2\bar\rho}{M_g^2}\delta = 0. \label{eq:exprdelta}
\end{equation}
The quantity $Q/\eta$, sometimes called $Y$ in the literature \cite{Amendola:2012ky,Amendola:2013qna,Konnig:2014dna}, represents deviations from Newton's constant in structure growth, and is effectively given in the subhorizon r\'{e}gime by $(h_1h_5)/(h_2h_3)$. We present the explicit forms of the $h_i$ coefficients in \cref{app:hcoeff}.

\subsection{Numerical solutions}

In this section we numerically solve for the background quantities and modified gravity parameters for one- and two-parameter bigravity models.\footnote{We focus on these simpler models to illustrate bigravity effects on growth. Current growth data are not able to significantly constrain these models, so we would not gain anything by adding more free parameters. The study of the full parameter space is left to future work.} We look in particular for potential observable signatures, as the growth data are currently not competitive with background data for these theories, although we expect future LSS experiments such as Euclid \cite{Laureijs:2011gra,Amendola:2012ys} to change this. The recipe is straightforward: using \cref{eq:brepyp} we can solve directly for $y(z)$, which is all we need to find solutions for $\eta(z,k)$ and $Q(z,k)$ using \cref{eq:expreta,eq:exprQ}. Finally these can be used, along with \cref{eq:brepH,eq:breprho}, to solve \cref{eq:exprdelta} numerically for $\delta(z,k)$ and hence for $f(z,k)$. We fit $f(z,k)$ to the parametrization $\Omega_m^\gamma$ in the redshift range $0<z<5$ unless stated otherwise.

The likelihoods for these models were analyzed in detail in Ref. \cite{Akrami:2012vf}, using the Union2.1 compilation of Type Ia supernovae \cite{Suzuki:2011hu}, Wilkinson Microwave Anisotropy Probe (WMAP) seven-year observations of the CMB \cite{Komatsu:2010fb}, and baryon acoustic oscillation (BAO) measurements from the galaxy surveys 2dFGRS, 6dFGS, SDSS and
WiggleZ \cite{Beutler:2011hx,Percival:2009xn,Blake:2011en}. We compute likelihoods based on growth data compiled in Ref. \cite{Macaulay:2013swa}, including growth histories from the 6dFGS \cite{Beutler:2012px}, LRG$_{200}$, LRG$_{60}$ \cite{Samushia:2011cs}, BOSS \cite{Tojeiro:2012rp}, WiggleZ \cite{Blake:2012pj} and VIPERS \cite{delaTorre:2013rpa} surveys.

Both the numerical solutions of background quantities and the likelihood computations are performed as in Ref. \cite{Akrami:2012vf}, where they are described in detail. Following Ref. \cite{Akrami:2012vf}, we will normalize the $\beta_i$ parameters to present-day Hubble rate, $H_0$, by defining
\begin{equation}
 B_i \equiv \frac{m^2}{H_0^2}\beta_i.
\end{equation}
Throughout, we will assume that the $g$-metric cosmological constant $B_0$ vanishes, as we are interested in the solutions which accelerate due to modified gravity effects.

\subsubsection{The minimal model}
\label{sec:b1}

We begin with the ``minimal'' model in which only $B_1$ is nonzero. This is the only single-$B_i$ theory which is in agreement with background observations \cite{Akrami:2012vf} (the other models also have theoretical viability issues \cite{Konnig:2013gxa}). Note, however, that the linear perturbations are unstable at early times until relatively recently, $z\sim0.5$ \cite{Konnig:2014xva}. This restricts the real-world applicability of the results presented herein, as the quasi-static approximation we employ will not be viable. Our results will hold for observations within the stable period,\footnote{Certainly our results for $Q$ and $\eta$ will hold; the growth rate $f$ may vary if the initial conditions for $\delta$ are significantly changed from what we assume herein.} and otherwise should be seen as an illustrative example.

\begin{figure}
\centering
\includegraphics[width=0.49\textwidth]{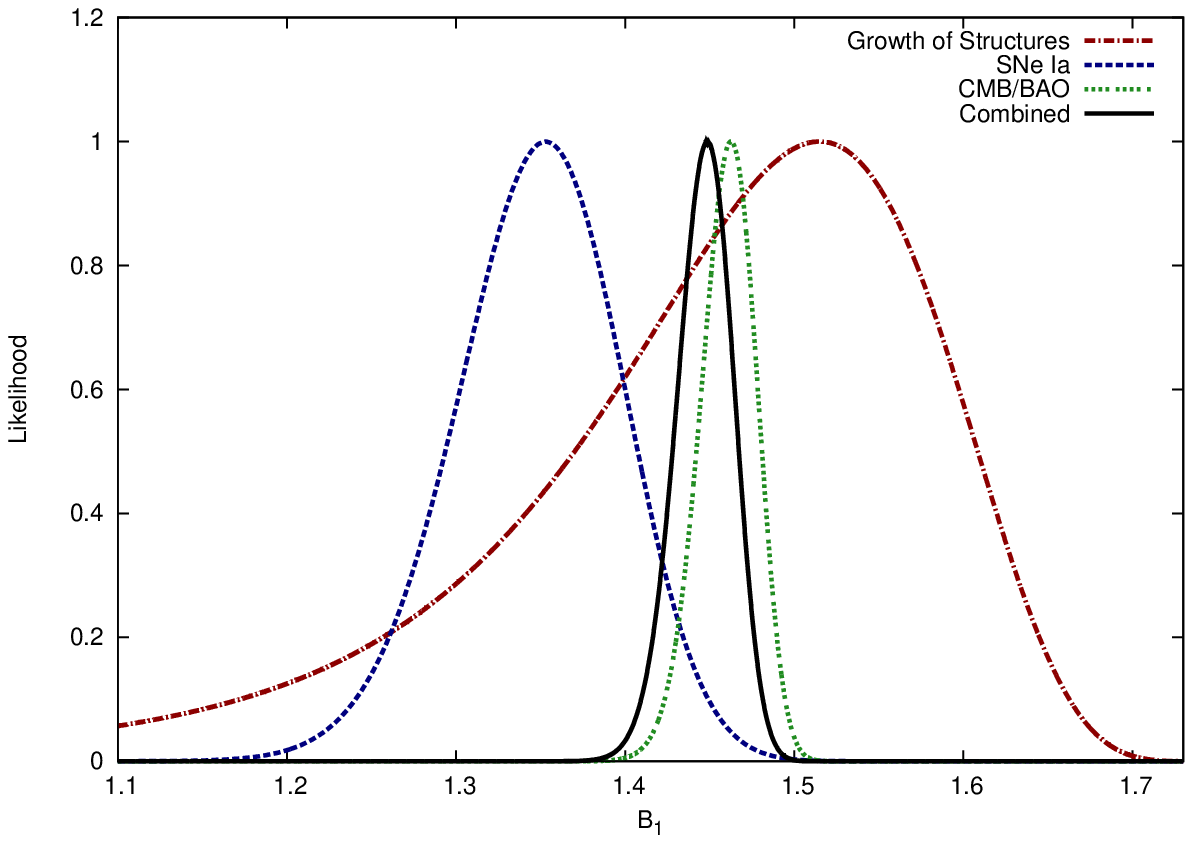}
\includegraphics[width=0.49\textwidth]{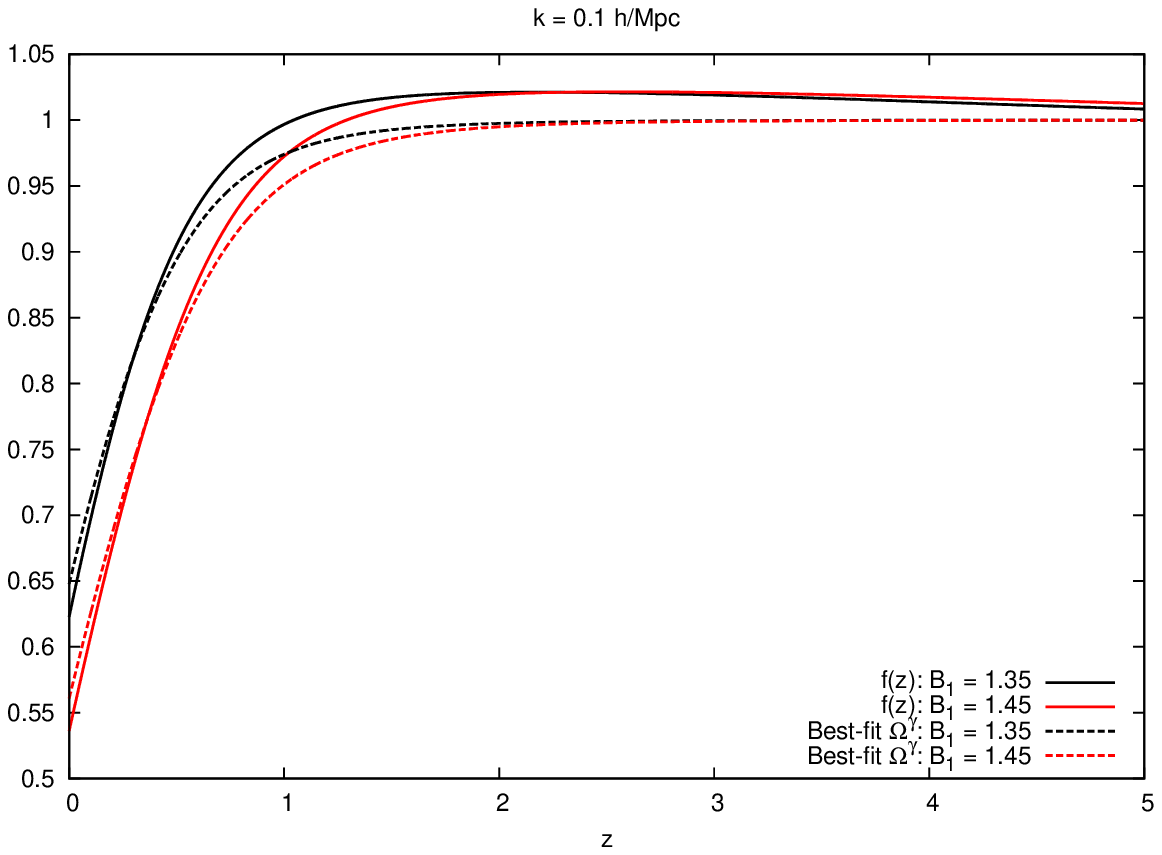}
\includegraphics[width=0.49\textwidth]{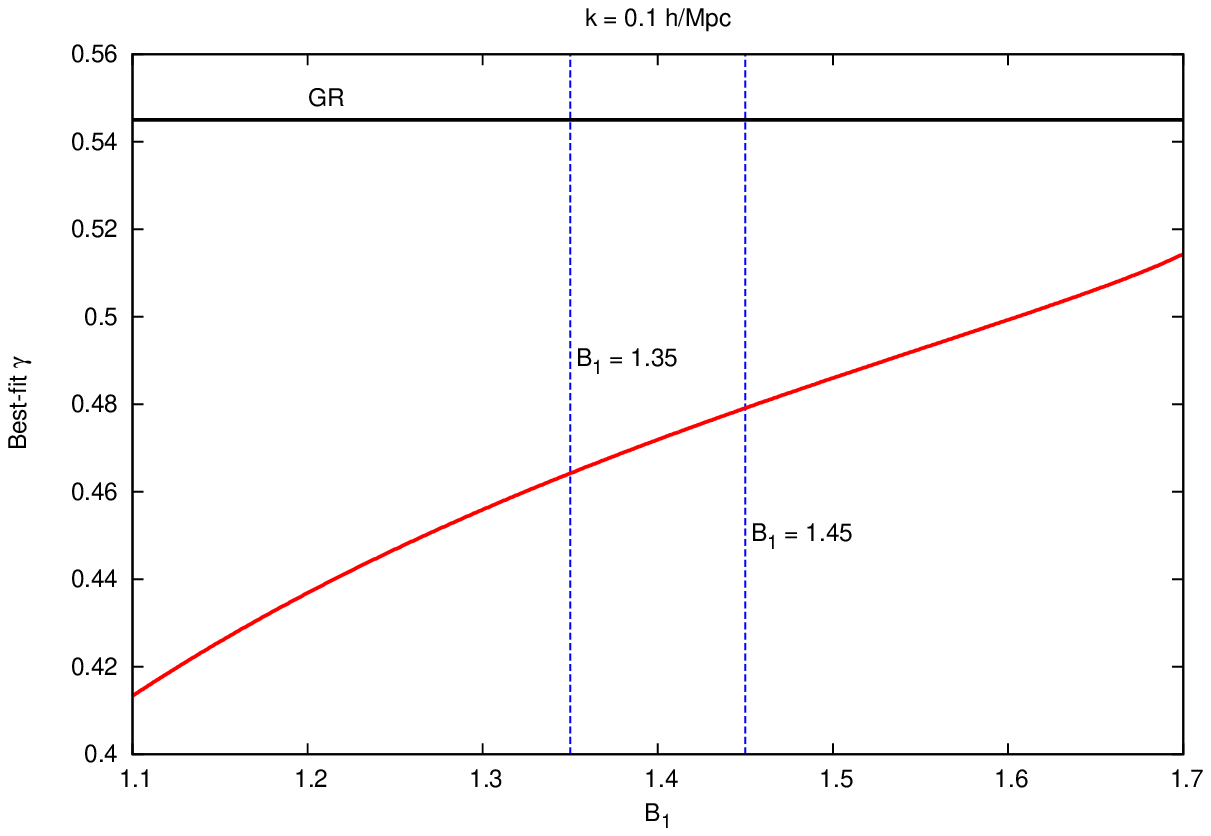}
\caption{\textbf{First panel:} The likelihood for $B_1$ in the $B_1$-only model from growth data (red), as well as background likelihoods for comparison. The fits for $B_1$ effectively depend only on the background data; the combined likelihood (black) is not noticeably changed by the addition of the growth data. \textbf{Second panel:} The growth rate, $f = d\ln\delta/d\ln a$, for the SNe best-fit parameter, $B_1=1.35$ (in black), and for the SNe/BAO/CMB combined best-fit parameter, $B_1=1.45$ (in red). The full growth rate (solid line) is plotted alongside the $\Omega^\gamma$ parametrization (dotted line) with best fits $\gamma=0.46$ and 0.48 for $B_1=1.35$ and 1.45, respectively. \textbf{Third panel:} The best-fit value of $\gamma$ as a function of $B_1$. For comparison, the GR prediction ($\gamma\approx0.545$) is plotted as a black horizontal line. The blue lines correspond to the best-fit values of $B_1$ from different background data sets. This is a prediction of a clear deviation from GR.}
\label{fig:b1-gamma}
\end{figure}

The likelihoods for $B_1$ are plotted in the first panel of \cref{fig:b1-gamma} based on supernovae, BAO/CMB, growth data, and all three combined, although the growth likelihood is so wide that it has a negligible effect on the combined likelihood. The point was raised in Ref. \cite{Konnig:2014dna} that the WMAP analysis is performed assuming a $\Lambda$CDM model and hence may not apply perfectly to these data. We will take an agnostic point of view on this and consider both the best-fit value of $B_1$ from supernovae alone ($B_1=1.3527\pm0.0497$) and from the combination of supernovae and CMB/BAO ($B_1=1.448\pm0.0168$).\footnote{These differ slightly from the best-fit $B_1=1.38\pm0.03$ reported by Ref. \cite{Konnig:2013gxa}, also based on the Union2.1 supernovae compilation.} The results do not change qualitatively with either choice.

The growth rate $f$ at $k=0.1$~$h$/Mpc is plotted in the second panel of \cref{fig:b1-gamma}, along with the parametrization $\Omega_m^\gamma$ with best fits $\gamma=0.46$ for $B_1=1.35$ and $\gamma=0.48$ for $B_1=1.45$. This is in agreement with the results of Ref. \cite{Konnig:2014dna}, who additionally found that $f(z)$ is fit much more closely by a redshift-dependent parametrization, $f \approx \Omega_m^{\gamma_0}\left(1+\frac{\gamma_1}{1+z}\right)$. In the third panel of \cref{fig:b1-gamma} we plot the best-fit value of $\gamma$ as a function of $B_1$. All values of $B_1$ consistent with background observations give a value of $\gamma$ that is far from the GR value (including $\Lambda$CDM and minimally-coupled quintessence models) of $\gamma\approx0.545$. While present observations of LSS are unable to easily distinguish this model from $\Lambda$CDM (cf. \cref{fig:b1-gamma}, first panel), the Euclid satellite expects to measure $\gamma$ within $0.02$ \cite{Laureijs:2011gra,Amendola:2012ys} and should easily be able to rule out either the minimal massive bigravity model or GR. Note that there is a caveat, in that we have calculated $\gamma$ by fitting over a redshift range ($0<z<5$) which includes the unstable period of this model's history ($z\gtrsim0.5$) during which linear theory breaks down. As emphasized above, these predictions should only be compared to data during the stable period. Therefore if this model does describe reality, Euclid may measure a different growth rate at higher redshifts; a nonlinear analysis is required to answer this with certainty.

\begin{figure}
\centering
\includegraphics[width=0.49\textwidth]{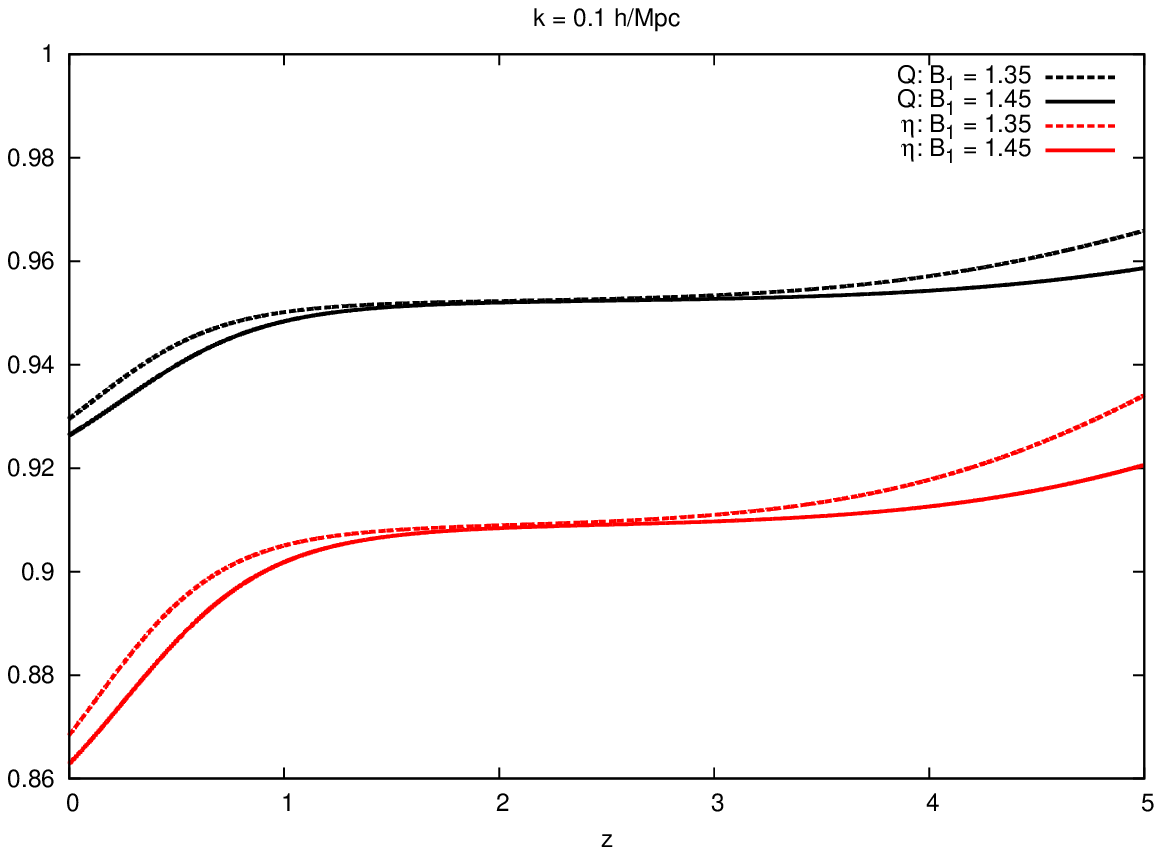}
\includegraphics[width=0.49\textwidth]{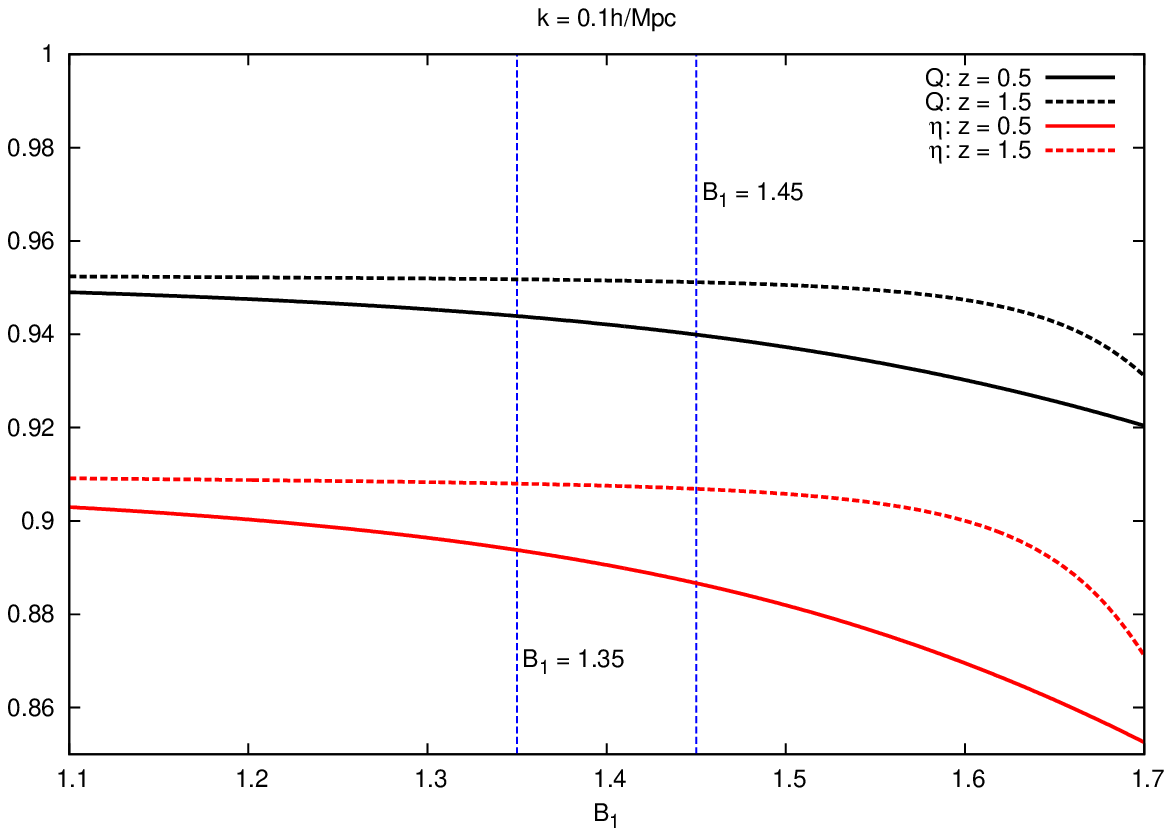}
\includegraphics[width=0.49\textwidth]{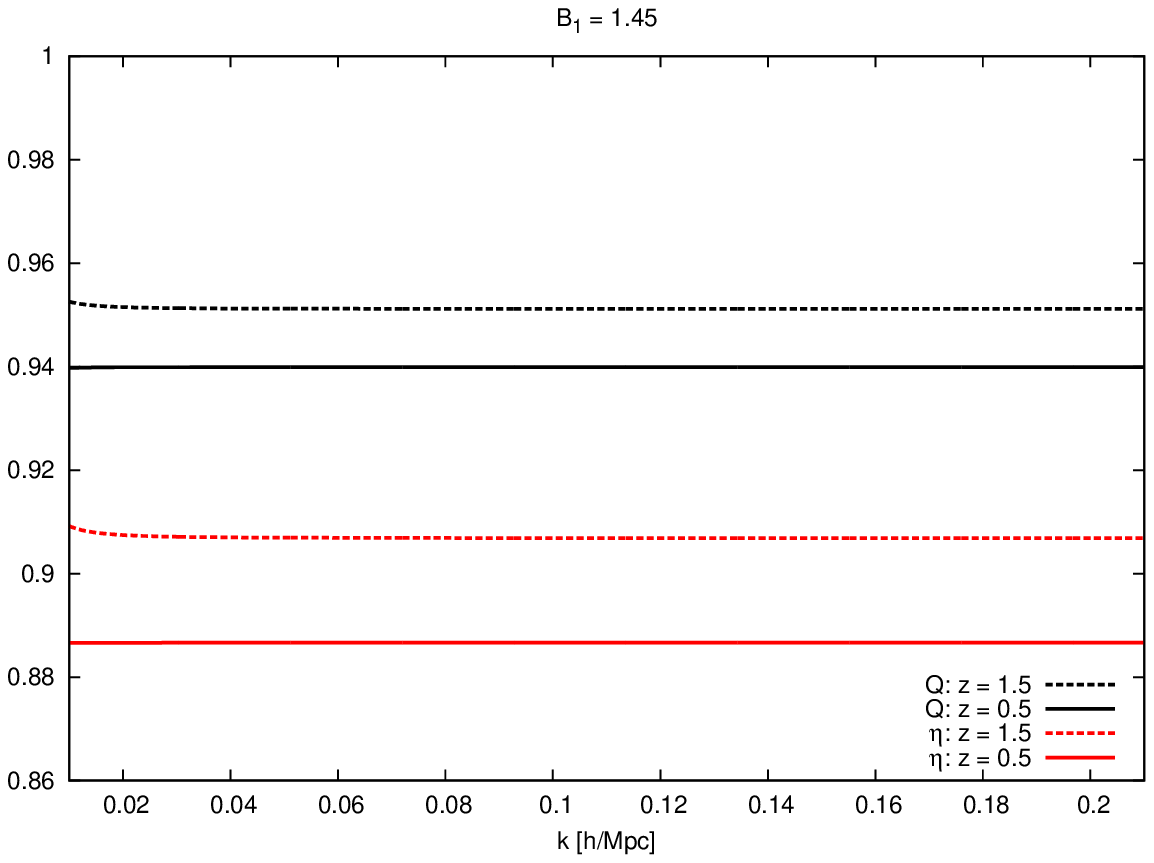}
\caption{The modified gravity parameters $Q$ (modification of Newton's constant) and $\eta$ (anisotropic stress) in the $B_1$-only model as a function of $z$, $B_1$, and $k$. They exhibit $\mathcal{O}(10^{-2})$--$\mathcal{O}(10^{-1})$ deviations from the GR prediction, which will be around the range of observability of a Euclid-like mission. They depend very weakly on $k$; consequently, more stringent constraints can be placed on them in a model-independent way by a future survey.}
\label{fig:b1-qe}
\end{figure}

We next look at the modified gravity parameters $\eta(z,k)$ and $Q(z,k)$. In \cref{fig:b1-qe} they are plotted with respect to $z$, $B_1$, and $k$, respectively, with the other two quantities fixed. $Q$ deviates from the GR value $Q=1$ by $\sim0.05$, while $\eta$ deviates from GR by up to $\sim0.15$. From the first panel of \cref{fig:b1-qe} we notice that $Q$ and $\eta$ lose their dependence on $B_1$ momentarily around $z\sim2.5$. This feature persists to other values of $B_1$ as well. Additionally, we can see from the third panel that $Q$ and $\eta$ only depend extremely weakly on $k$ in the linear subhorizon r\'egime. Future structure experiments like Euclid will be able to constrain $Q$ and $\eta$ more tightly in a model-independent way because they are effectively scale-independent; in particular, because of scale independence they are expected to be able to measure $\eta$ within 10\% \cite{Amendola:2013qna}, which would bring this minimal model to the cusp of observability.

\subsubsection{Two-parameter models}

At the background level, four models with two nonzero $B_i$ parameters provide good fits to the data: $B_1$--$B_2$, $B_1$--$B_3$, $B_1$--$B_4$, and $B_2$--$B_3$ \cite{Akrami:2012vf}. Even though these models all possess a two-dimensional parameter space, only an effectively one-dimensional subspace matches the background data (cf. figures 7 and 8 of Ref. \cite{Akrami:2012vf} and figure 4 of Ref. \cite{Konnig:2013gxa}). We will restrict ourselves to those subspaces by fixing one $B_i$ parameter in terms of another, usually $B_1$. We do this by identifying the effective present-day dark energy density, $\Omega_\Lambda^\mathrm{eff}$, from the Friedmann equation (\ref{eq:friedmann_g}):
\begin{equation}
\Omega_\Lambda^\mathrm{eff} \equiv B_1y_0 + B_2y_0^2 + \frac{1}{3}B_3y_0^3, \label{eq:omegalambdadef}
\end{equation}
and plugging that into the quartic equation for $y$ (\ref{eq:quartic}), evaluated at the present day ($y=y_0$) and using $\Omega_{m,0} + \Omega_\Lambda^\mathrm{eff}=1$. This procedure fixes one $B_i$ parameter in terms of the other and $\Omega_\Lambda^\mathrm{eff}$. The value of $\Omega_\Lambda^\mathrm{eff}$ can be determined by fitting to the data, as was done in Ref. \cite{Akrami:2012vf}. However, for the $B_1$--$B_i$ models we can also simply take the limit in which only $B_1$ is nonzero, recovering the single-parameter model discussed above; by then using the SNe/CMB/BAO combined best-fit value $B_1=1.448$ we find $\Omega_\Lambda^\mathrm{eff} = 0.699$.

A detailed study of the conditions for a viable background was undertaken in Ref. \cite{Konnig:2013gxa}. There are two results that are particularly relevant for the present study and bear mentioning. First, a viable model requires $B_1\geq0$. Second, the two-parameter models are all (with one exception, which we discuss below) ``finite branch" models, in which $y$ evolves from 0 at $z=\infty$ to a finite value $y_c$ at late times, which can be determined from \cref{eq:breprho} by setting $\rho=0$ at $y=y_c$. Consequently, the present-day value of $y_0$, which is generally simple to calculate, must always be smaller than $y_c$ (or above $y_c$ in a model with an infinite branch). In the $B_1$--$B_3$ and $B_1$--$B_4$ models this will rule out certain regions of the parameter space \textit{a priori}.

All of the two-parameter models except for the infinite-branch $B_1$--$B_4$ model (so-called ``infinite-branch bigravity'') suffer from the early-time instability found in Ref. \cite{Konnig:2014xva}; therefore caution should be used when applying the results for any of the other models in this section to real-world data. We emphasize again that our quasistatic approximation is only valid at low redshifts, and that moreover the growth rate should be re-calculated using whatever initial conditions the earlier period ends with. (The predictions for $Q$ and $\eta$ do not depend on solving any differential equation and therefore apply without change.) Modulo this caveat, we present the quasistatic results for the unstable models as proofs of concept, as examples of how to apply our methods. The infinite-branch bigravity predictions, presented at the end of this section, can be straightforwardly applied to data.

We evaluate all quantities at $k=0.1$~$h$/Mpc. The modified gravity parameters in all of the two-parameter models we study depend extremely weakly on $k$, as in the $B_1$-only model.

As we have already mentioned, in these models the two $B_i$ parameters are highly degenerate at the background level. One of the main goals of this section is to see whether observations of LSS have the potential to break this degeneracy.

\paragraph{$\mathbf{B_1}$--$\mathbf{B_2}$:} 

The $B_1$--$B_2$ models which fit the background data \cite{Akrami:2012vf} live in the parameter subspace
\begin{equation}
B_2=\frac{-B_1^2 + 9 \Omega_\Lambda^\mathrm{eff} - \sqrt{B_1^4 + 9 B_1^2 \Omega_\Lambda^\mathrm{eff}}}{9\Omega_\Lambda^\mathrm{eff}}, \label{eq:b2b1rel}
\end{equation}
with $\Omega_\Lambda^\mathrm{eff} \approx 0.7$. This line has a slight thickness because we must rely on observations to fit $\Omega_\Lambda^\mathrm{eff}$. We can subsequently determine $y_0$ from \cref{eq:omegalambdadef}.

\begin{figure}
\centering
\includegraphics[width=0.49\textwidth]{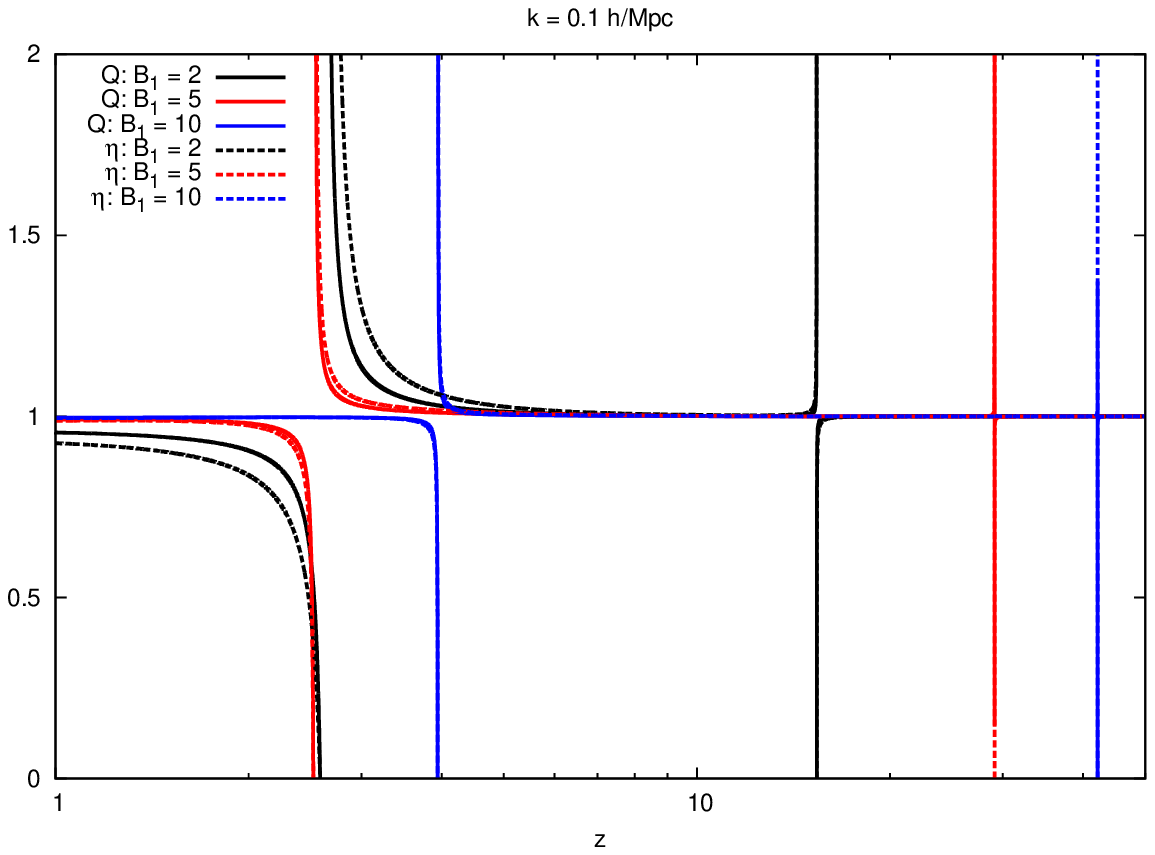}
\includegraphics[width=0.49\textwidth]{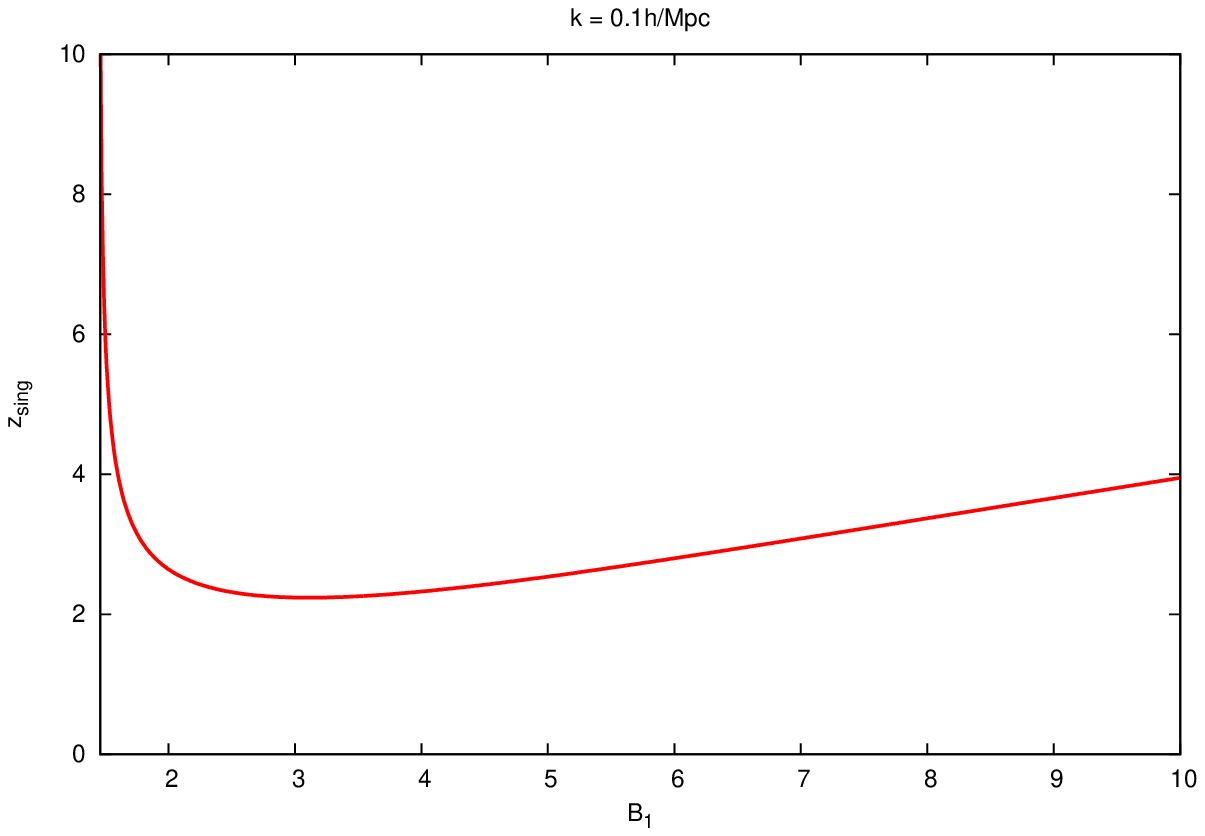}
\caption{\textbf{First panel:} $Q$ and $\eta$ for a few parameter values in the instability range of the $B_1$--$B_2$ model. Generally there are two poles, each of short duration. Note that the redshift at which the pole occurs depends only on $B_1$ and $B_2$ and not on the initial conditions for the perturbations. \textbf{Second panel:} The redshift of the most recent pole in $Q$ (the pole in $\eta$ occurs nearly simultaneously) in terms of $B_1>1.448$. This parameter range corresponds to $B_2<0$, cf. \cref{eq:b2b1rel}. The minimum is at $(B_1,B_2,z)=(3.11,-2.51,2.24)$.}
\label{fig:zsing}
\end{figure}

This model possesses an instability when $B_2<0$.\footnote{A similar singular evolution of linear perturbations in a smooth background has been observed in the cosmology of Gauss-Bonnet gravity \cite{Kawai:1999pw,Koivisto:2006xf}. This instability is, however, different from the early-time instabilities discussed above, as those instabilities do not arise in the quasistatic limit which we are now taking.} This is not entirely unexpected: the $B_2$ term is the coefficient of the quadratic interaction, and so a negative $B_2$ might lead to a tachyonic instability. However, the instability of the $B_1$--$B_2$ model is somewhat unusual: in the subhorizon limit, $Q$ and $\eta$ develop poles, but they only diverge during a brief period around a fixed redshift, as shown in the first panel of \cref{fig:zsing}, regardless of wavenumber or initial conditions. We can find these poles using the expressions in \cref{app:hcoeff}. The exact solutions are unwieldy and not enlightening, but there are three notable features. First, as mentioned, the instability only occurs for $B_2<0$. (When $B_2<0$, these poles occur when $y$ is negative, which is not physical.) Second, the instability develops at high redshifts, $z>2$. The redshift of the latest pole (which can be solved for by taking the limit $k/H\to\infty$) is plotted as a function of $B_1$ in the second panel of \cref{fig:zsing}. As a result, measurements of $Q$ and $\eta$ at $z\lesssim2$ would generally not see divergent values. However, such measurements would see the main instabilities at much lower redshifts \cite{Konnig:2014xva}. Finally, the most recent pole occurs at $y=0$ for $B_2=0$ ($B_1=\sqrt{3\Omega_\Lambda^\mathrm{eff}} \approx 1.45$), and approaches $y=y_0/2$ as $B_2\to-\infty$ ($B_1\to\infty$).

\begin{figure}
\centering
\includegraphics[width=0.49\textwidth]{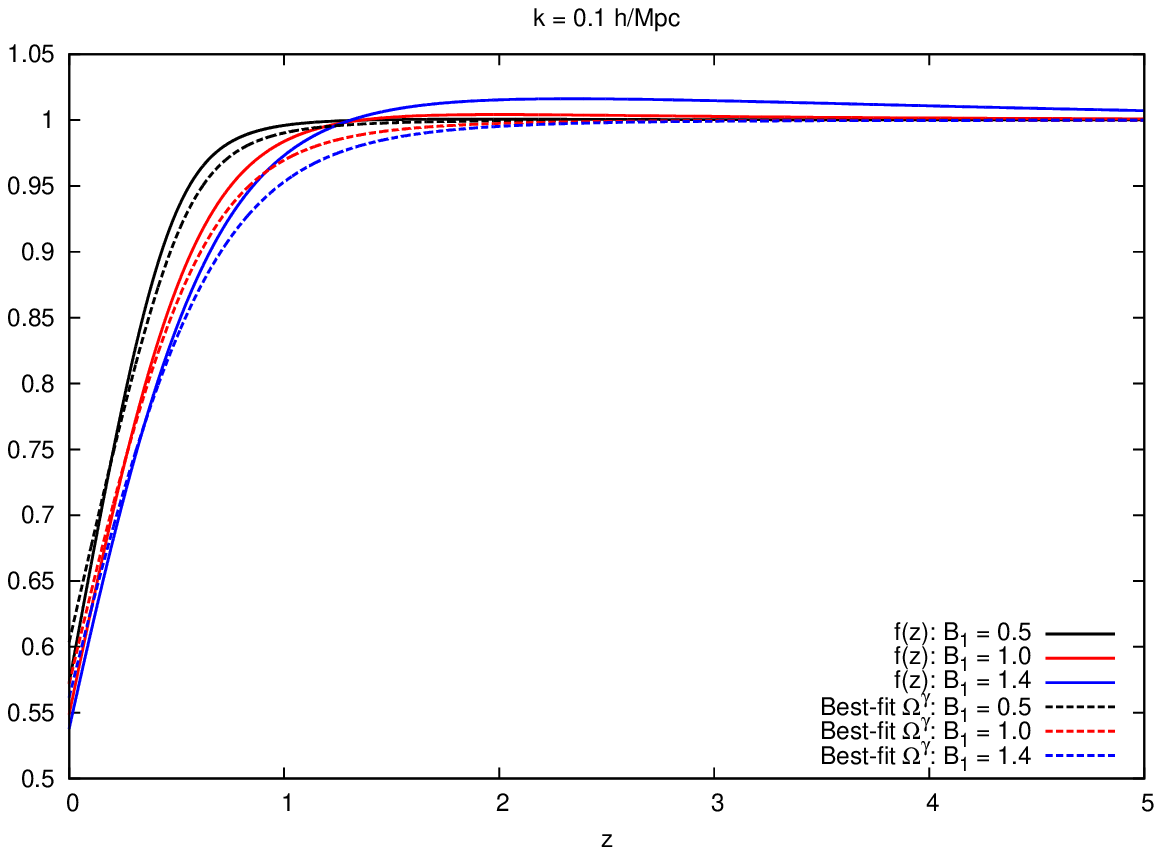}
\includegraphics[width=0.49\textwidth]{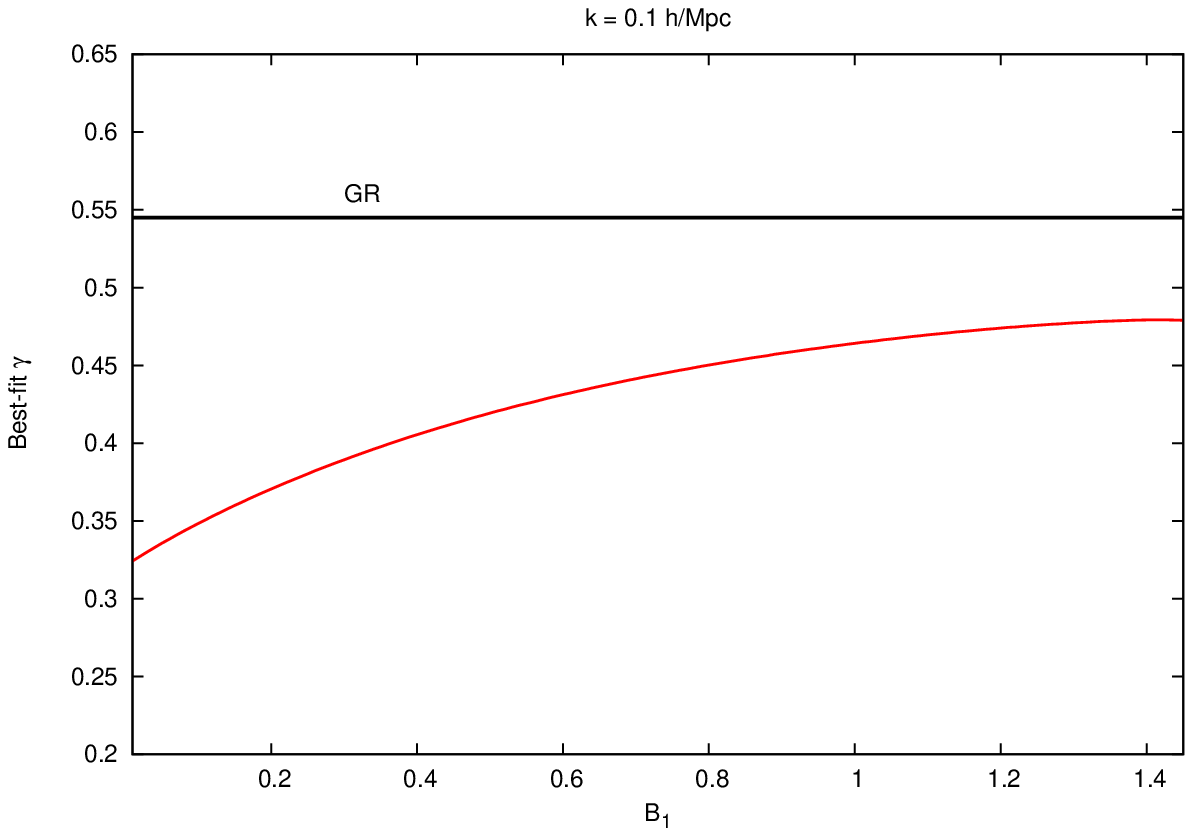}
\includegraphics[width=0.49\textwidth]{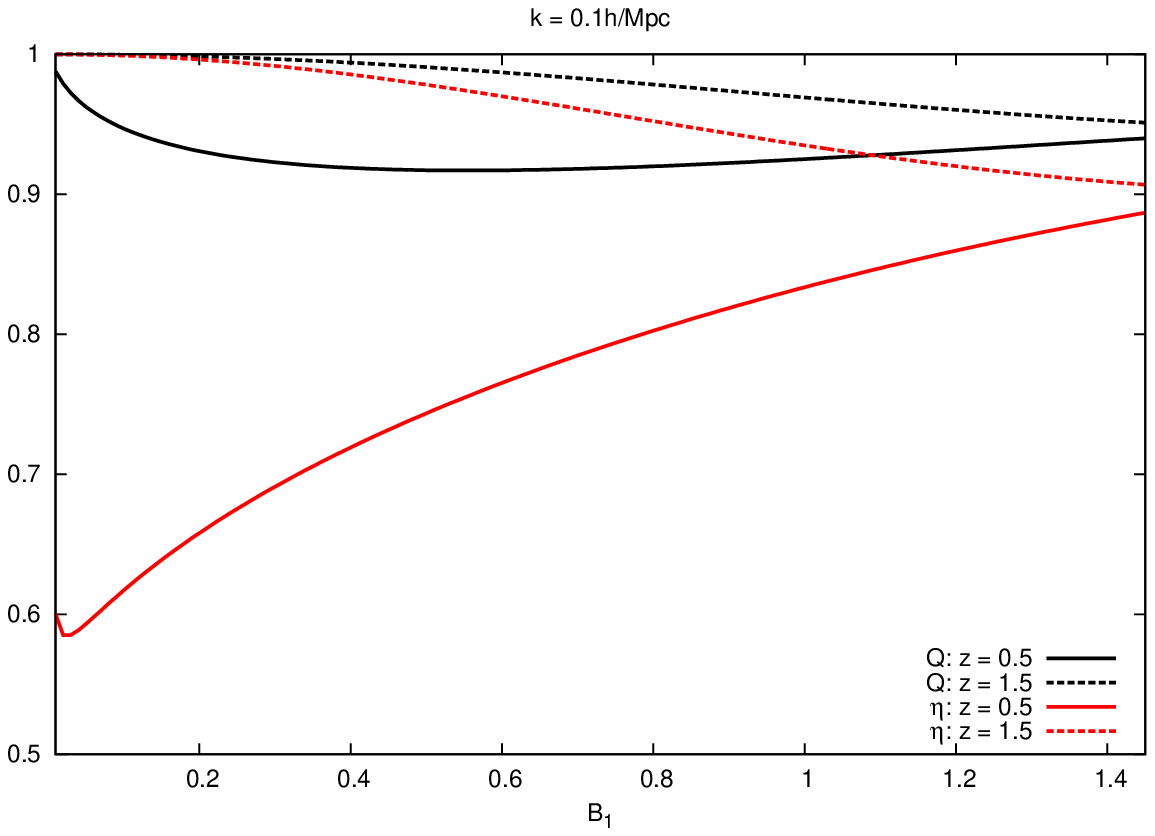}
\includegraphics[width=0.49\textwidth]{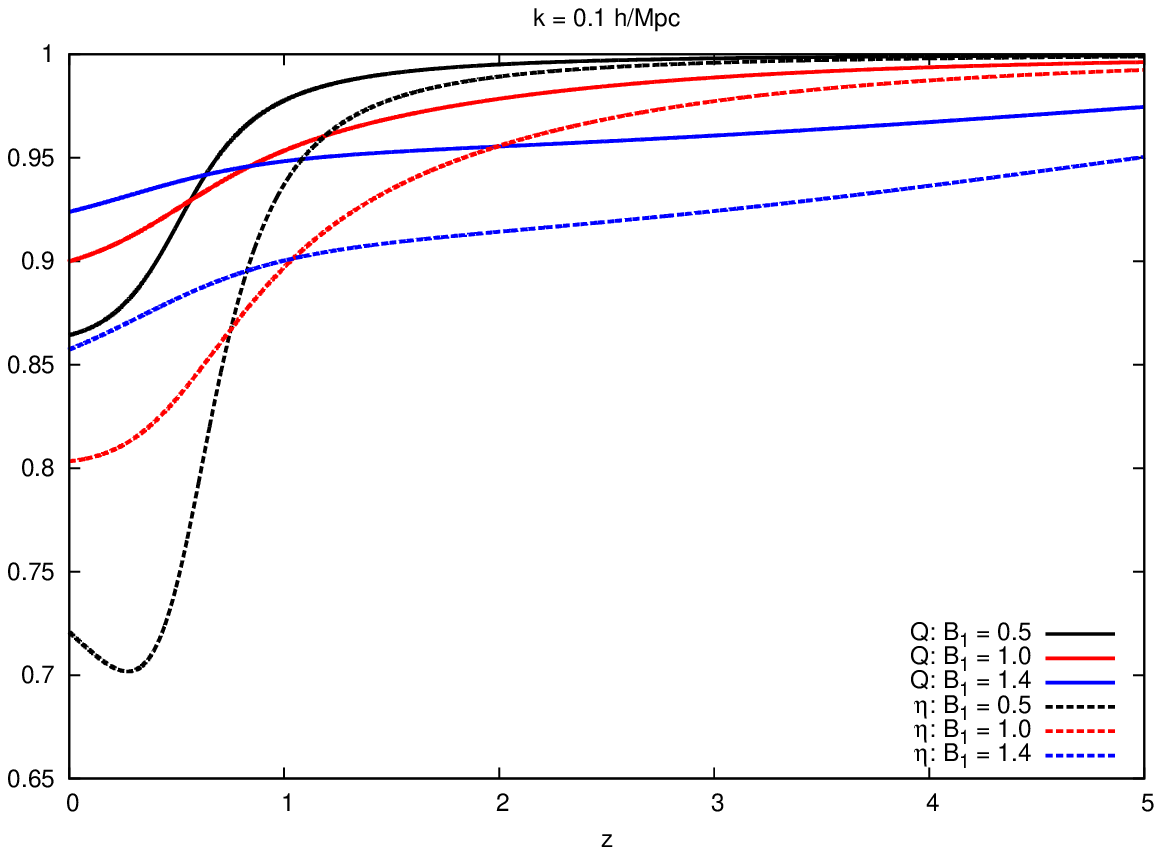}
\caption{Results for the $B_1$--$B_2$ model, with $\Omega_\Lambda^\mathrm{eff}=0.699$ and $B_1\leq1.448$. The significant deviations from GR in $\gamma$ and $\eta(z\sim0.5)$ should be observable by a Euclid-like experiment. Moreover, they have the potential to break the degeneracy between $B_1$ and $B_2$ when fitting to background observations.}
\label{fig:b1-b2}
\end{figure}

This particular instability is avoided if we restrict ourselves to the range $0<B_1\leq1.448$, for which $B_2>0$. Some typical results for this region of parameter space are plotted in \cref{fig:b1-b2}. The first panel plots $f(z)$ and the best-fit $\Omega_m^\gamma$ parametrization for selected values of $B_1$ (with $B_2$ given by \cref{eq:b2b1rel}), while the second panel shows the best-fit value of $\gamma$ as a function of $B_1$. For smaller values of $B_1$ this parametrization fits $f$ well, more so than in the $B_1$-only model discussed in \cref{sec:b1} (which is the $B_1=1.448$ limit of this model). We find that $\gamma$ is always well below the GR value of $\gamma\approx0.545$, especially at low $B_1$.

In the final two panels we plot $Q$ and $\eta$, both in terms of $B_1$ at fixed $z$ and in terms of $z$ at fixed $B_1$. In comparison to the $B_1$-only model (the very right side of the third panel), lowering $B_1$ tends to make these parameters more GR-like, except for $\eta$ evaluated at late times ($z\sim0.5$), which dips as low as $\eta\sim0.6$. Because these quantities are all $k$-independent in the linear, subhorizon r\'egime, future LSS experiments like Euclid would be able to measure $\eta$ at these redshifts to within about 10\% \cite{Amendola:2013qna} and thus effectively distinguish between $\Lambda$CDM and significant portions of the parameter space of the $B_1$--$B_2$ model, testing the theory and breaking the background-level degeneracy.

\paragraph{$\mathbf{B_1}$--$\mathbf{B_3}$:} 

The $B_1$--$B_3$ models which are consistent with the background data \cite{Akrami:2012vf} lie along a one-dimensional parameter space (up to a slight thickness) given by
\begin{equation}
B_3 = \frac{-32 B_1^3 + 81 B_1 \Omega_\Lambda^\mathrm{eff} \pm \sqrt{\left(8 B_1^2-27 \Omega_\Lambda^\mathrm{eff}\right)^2 \left(16 B_1^2+27 \Omega_\Lambda^\mathrm{eff}\right)}}{243 (\Omega_\Lambda^\mathrm{eff})^2}, \label{eq:b3b1rel}
\end{equation}
with $\Omega_\Lambda^\mathrm{eff} \approx 0.7$. There is a subtlety here: the physical branch is constructed in a piecewise fashion, taking the $+$ root for $B_1<(3/2)^{3/2}\sqrt{\Omega_\Lambda^\mathrm{eff}}= 1.536$ and the $-$ root otherwise \cite{Konnig:2013gxa}. We solve for $y(z)$ using the initial condition,\footnote{There is also a positive root, but this is not physical. When $B_3<0$, that root yields $y_0<0$. When $B_3>0$, which is only the case for a small range of parameters, then the positive root of $y_0$ is greater than the far-future value $y_c$ and hence is also not physical.} derived from \cref{eq:quartic},
\begin{equation}
y_0 = \frac{1-\sqrt{1-\frac{4}{3}B_1B_3}}{2B_3}.
\end{equation}
As discussed at the beginning of this section, $y_0$ should not be larger than the value of $y$ in the far future, $y_c$. Demanding this, we find a maximum allowed value for $B_3$,\footnote{Note, per \cref{eq:b3b1rel}, that this is equivalent to simply imposing $\Omega_\Lambda^\mathrm{eff}<1$, which must be true since we have chosen a spatially-flat universe \textit{a priori}.}
\begin{equation}
B_3 < \frac{1}{243}\left(-32B_1^3 + 81B_1+\sqrt{\left(16B_1^2+27\right)\left(8B_1^2-27\right)^2}\right).
\end{equation}
For $\Omega_\Lambda^\mathrm{eff}=0.699$ this implies we need to restrict ourselves to $B_1>1.055$. This sort of bound is to be expected: we know that the $B_3$-only model is a poor fit to the data \cite{Akrami:2012vf}, so we cannot continue to get viable cosmologies the entire way through the $B_1\to0$ limit of the $B_1$--$B_3$ model.

\begin{figure}
\centering
\includegraphics[width=0.49\textwidth]{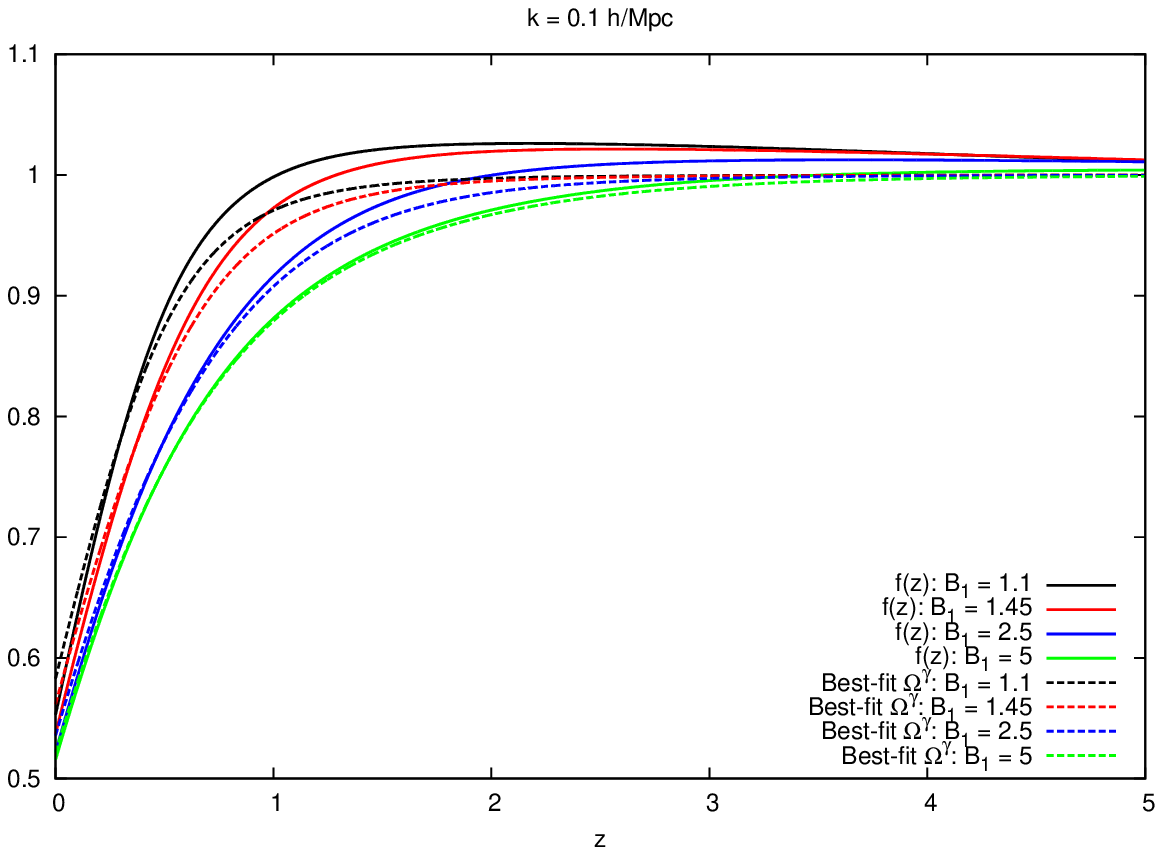}
\includegraphics[width=0.49\textwidth]{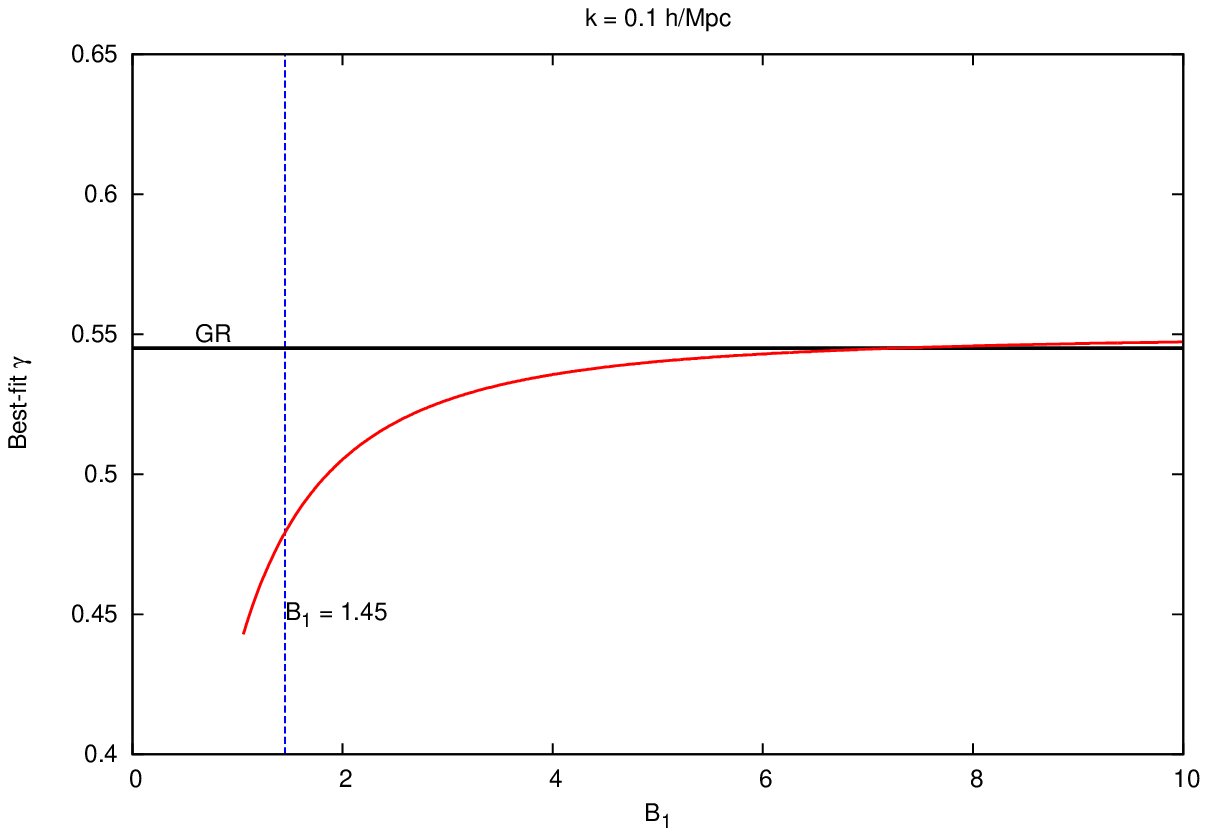}
\includegraphics[width=0.49\textwidth]{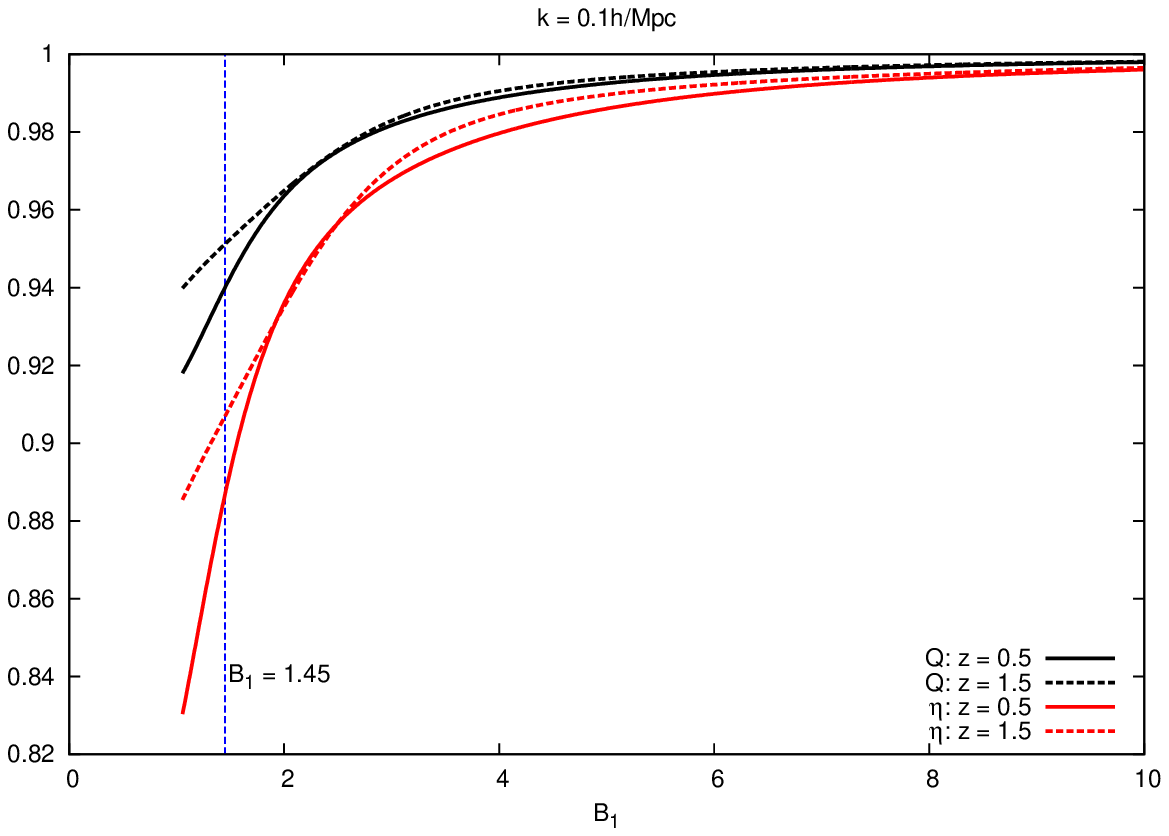}
\includegraphics[width=0.49\textwidth]{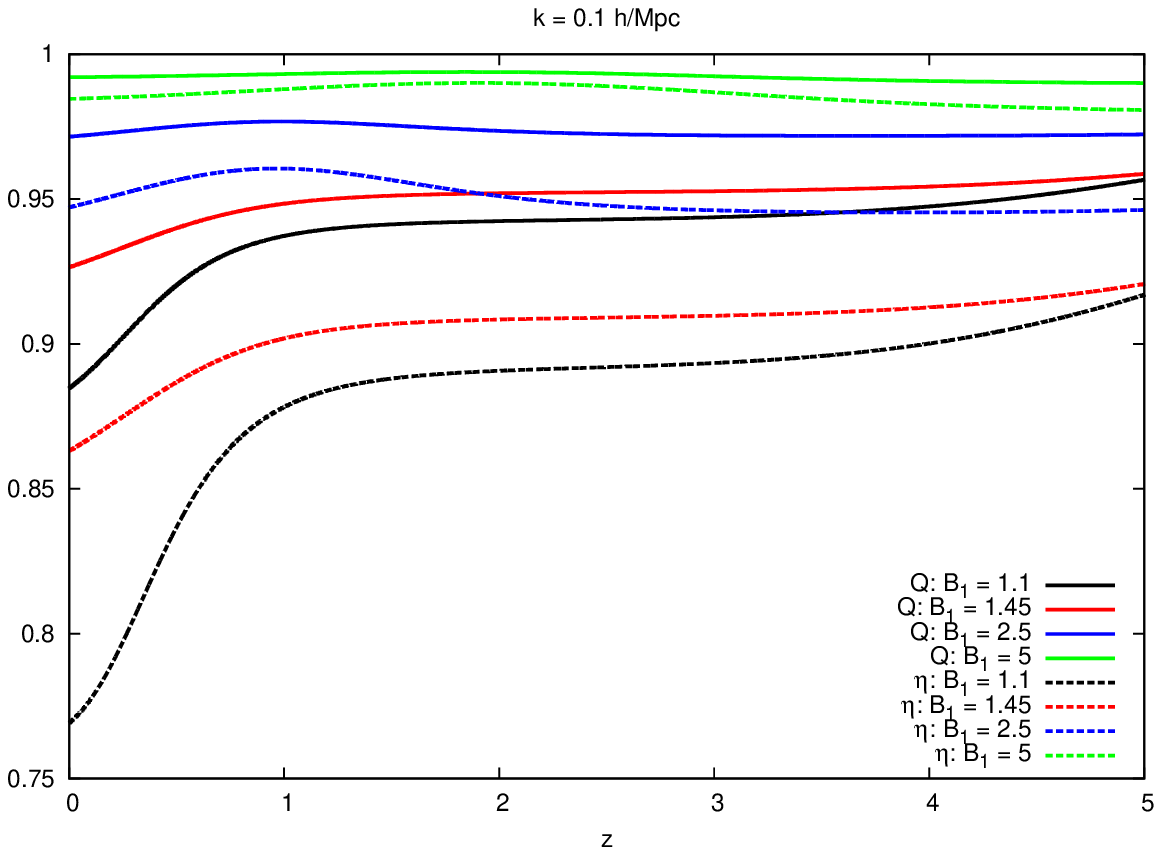}
\caption{Results for the $B_1$--$B_3$ model, with $\Omega_\Lambda^\mathrm{eff}=0.699$ and $B_1>1.055$. While $B_1$ and $B_3$ are degenerate in the background (along a line given by \cref{eq:b3b1rel}), perturbations clearly can break this degeneracy, with significant deviations at low values of $B_1$ (small $|B_3|$), and GR-like behavior at large values of $B_1$ (large, negative $B_3$).}
\label{fig:b1-b3}
\end{figure}

We plot the results for the $B_1$--$B_3$ model in \cref{fig:b1-b3}. These display the tendency, which we will also see in the $B_1$--$B_4$ model, that large $|B_i|$ values lead to modified gravity parameters that are closer to GR. For example, $\gamma$ can be as low as $\gamma\approx0.45$ for the lowest allowed value of $B_1$, but by $B_1\sim3$ it is practically indistinguishable from the GR value, assuming a Euclid-like precision of $\sim0.02$ on $\gamma$ \cite{Laureijs:2011gra}. Again we note that this value of $\gamma$ has been obtained assuming $0<z<5$, which is not a valid range for observations because of the early-time instability. For lower values of $B_1$, current growth data (see, e.g., Ref. \cite{Macaulay:2013swa}) are not sufficient to significantly constrain the parameter space, but these non-GR values of $\gamma$ and $\eta$ should be well within Euclid's window.

\paragraph{$\mathbf{B_1}$--$\mathbf{B_4}$:}

This model comprises the lowest-order $B_1$ term in conjunction with a cosmological constant for the $f$ metric, $B_4$.\footnote{In the singly-coupled version of massive bigravity we are studying, matter loops only contribute to the $g$-metric cosmological constant, $B_0$.} Note that $B_4$ does not contribute directly to the Friedmann equation (\ref{eq:friedmann_g}), but only affects the dynamics through its effect on the evolution of $y$.

The $B_1$--$B_4$ model has two viable solutions for $y(z)$: a finite branch with $0<y<y_c$, and an infinite branch with $y_c<y<\infty$. The infinite-branch model is the only two-parameter bimetric model which is linearly stable at all times \cite{Konnig:2014xva}. Therefore this \textit{infinite-branch bigravity} should be considered the most viable bimetric massive gravity theory to date. In this section we will elucidate its predictions for subhorizon structure formation.

As discussed in the beginning of this section, $y_c$ is the value of $y$ in the asymptotic future and can be calculated by setting $\rho=0$ in \cref{eq:breprho}. We will consider the two branches separately.

For a given $\Omega_\Lambda^\mathrm{eff}$, $B_4$ is related to $B_1$ by
\begin{equation}
B_4 = \frac{3\Omega_\Lambda^\mathrm{eff}B_1^2 - B_1^4}{(\Omega_\Lambda^\mathrm{eff})^3}, \label{eq:b1-b4}
\end{equation}
while $y_0$ is given by
\begin{equation}
 y_0 = \frac{\Omega_\Lambda^\mathrm{eff}}{B_1}.
\end{equation}
The finite-branch (infinite-branch) condition $y_0<y_c$ ($y_0>y_c$) imposes
\begin{equation}
B_4 > 2B_1\qquad (B_4 < 2B_1). \label{eq:b1-b4-branch}
\end{equation}
The combination of \cref{eq:b1-b4,eq:b1-b4-branch} places a constraint on the allowed range of $B_1$, as in the $B_1$--$B_3$ model, which depends on the best-fit value of $\Omega_\Lambda^\mathrm{eff}$. The $B_1$-only model ($B_4=0$, $B_1>0$) is on the finite branch, so that on that branch we can use $\Omega_\Lambda^\mathrm{eff}=0.699$ as we did in the other models. This implies $B_1>1.244$ for the finite branch. On the infinite branch, SNe observations are best fit by $\Omega_\Lambda^\mathrm{eff}=0.84$ \cite{Konnig:2013gxa}; consequently we restrict ourselves to $B_1<0.529$ for the infinite branch.

\begin{figure}
\centering
\includegraphics[width=0.49\textwidth]{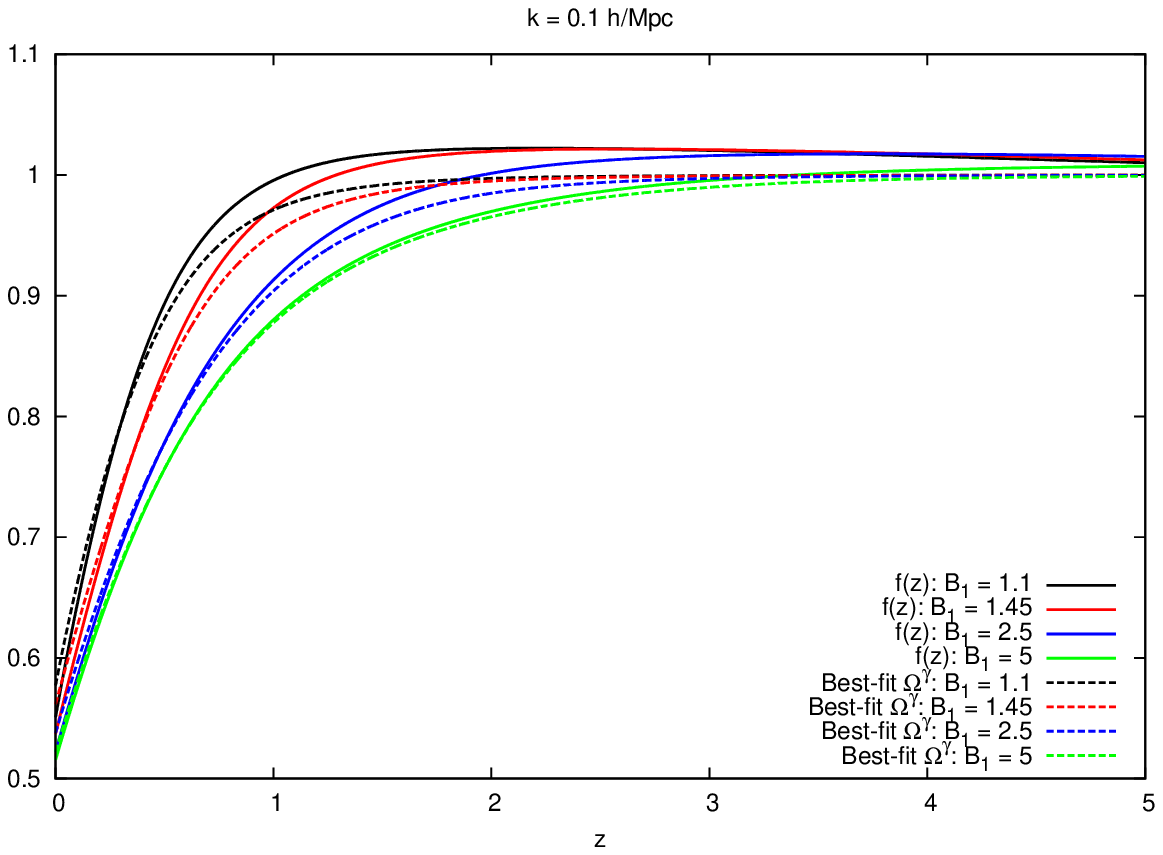}
\includegraphics[width=0.49\textwidth]{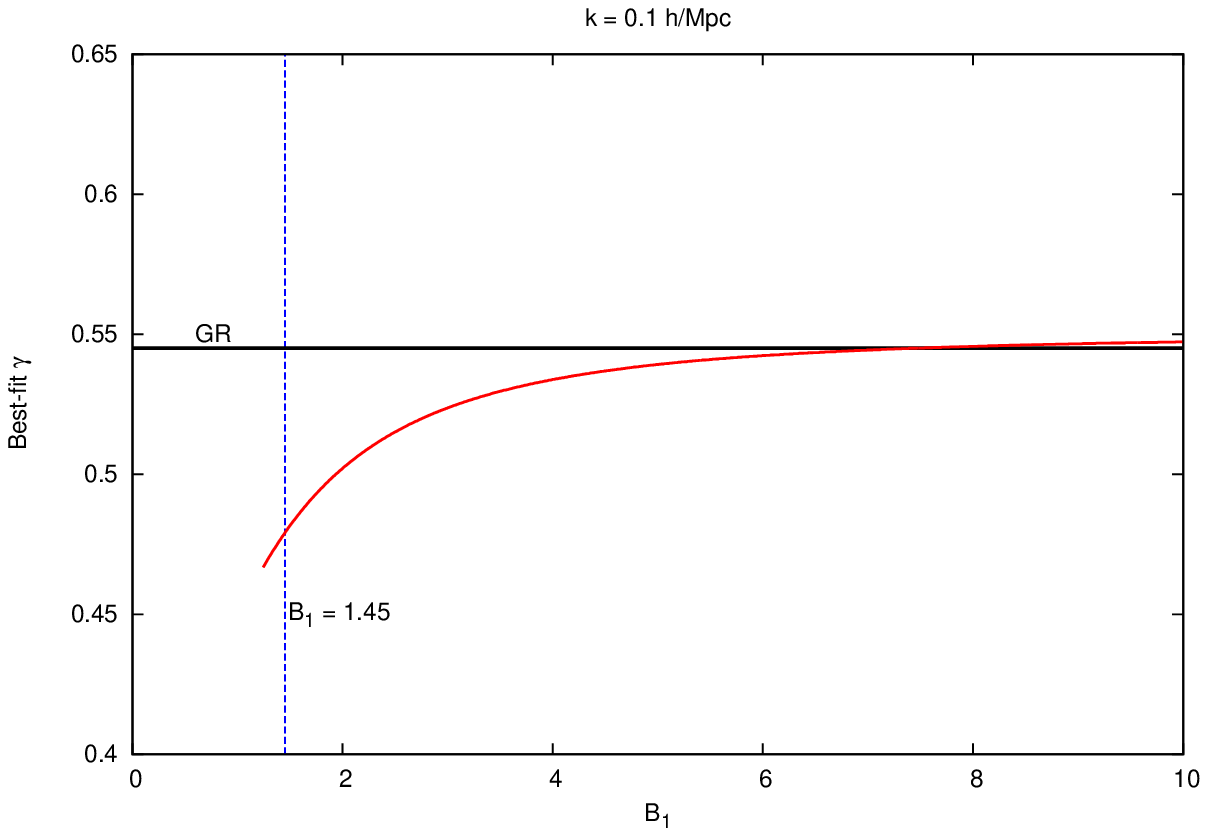}
\includegraphics[width=0.49\textwidth]{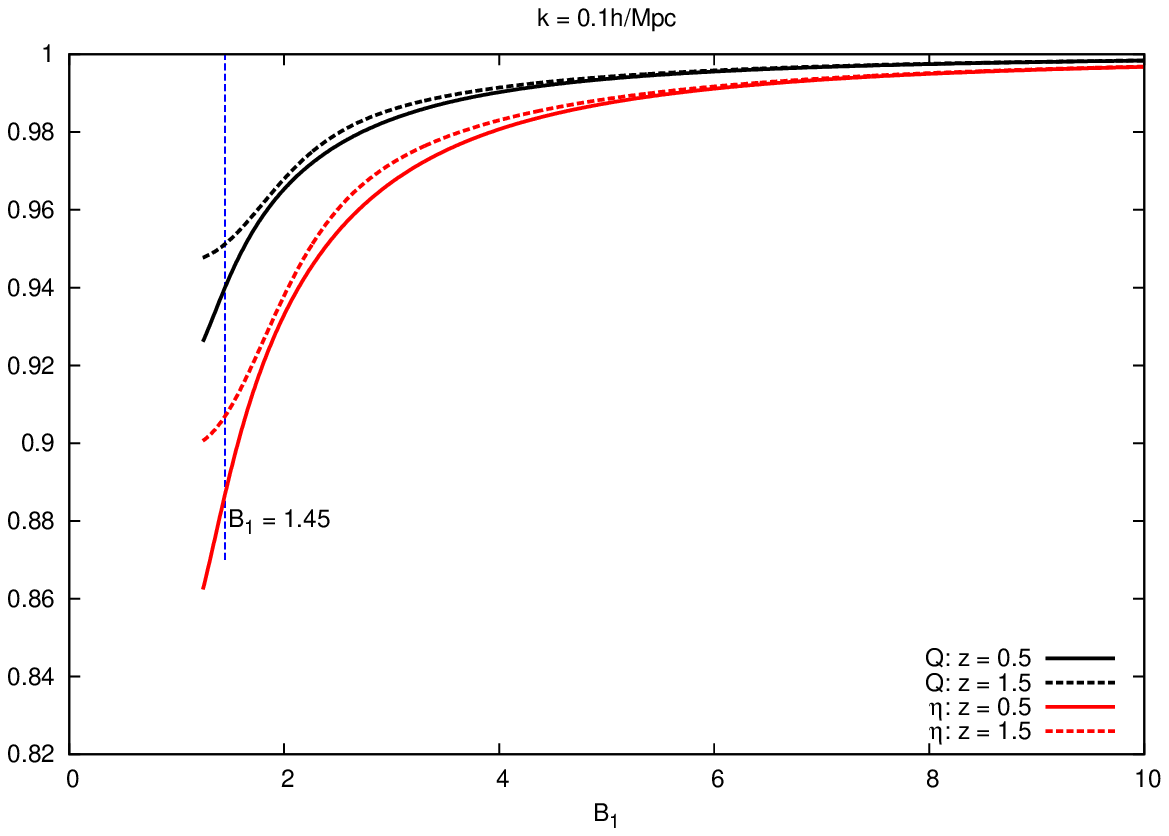}
\includegraphics[width=0.49\textwidth]{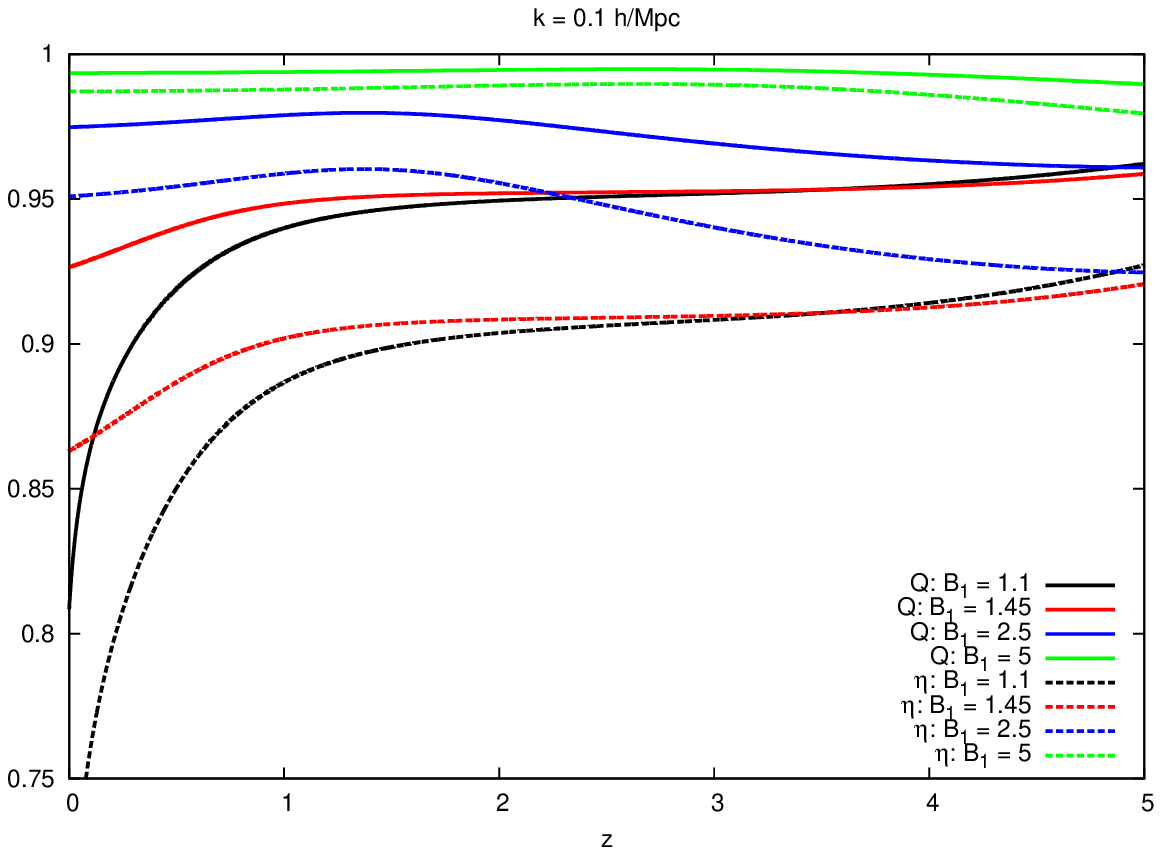}
\caption{Results for the $B_1$--$B_4$ model on the finite branch, with $\Omega_\Lambda^\mathrm{eff}=0.699$ and $B_1>1.244$. The behavior is quite similar to that of the $B_1$--$B_3$ model, plotted in \cref{fig:b1-b3}.}
\label{fig:b1-b4-finite}
\end{figure}

We plot the results for the finite branch in \cref{fig:b1-b4-finite}. Qualitatively, this model predicts subhorizon behavior quite similar to that of the $B_1$--$B_3$ model, discussed above and plotted in \cref{fig:b1-b3}.

\begin{figure}
\centering
\includegraphics[width=0.49\textwidth]{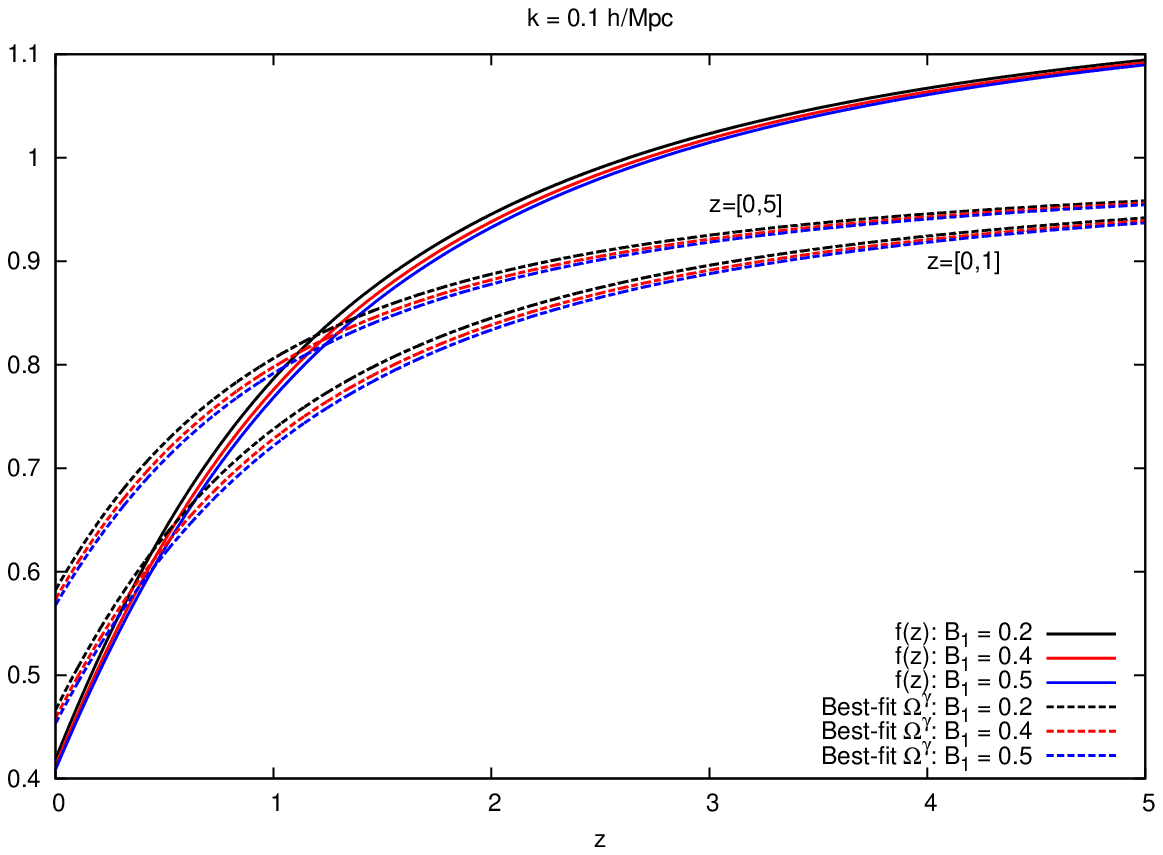}
\includegraphics[width=0.49\textwidth]{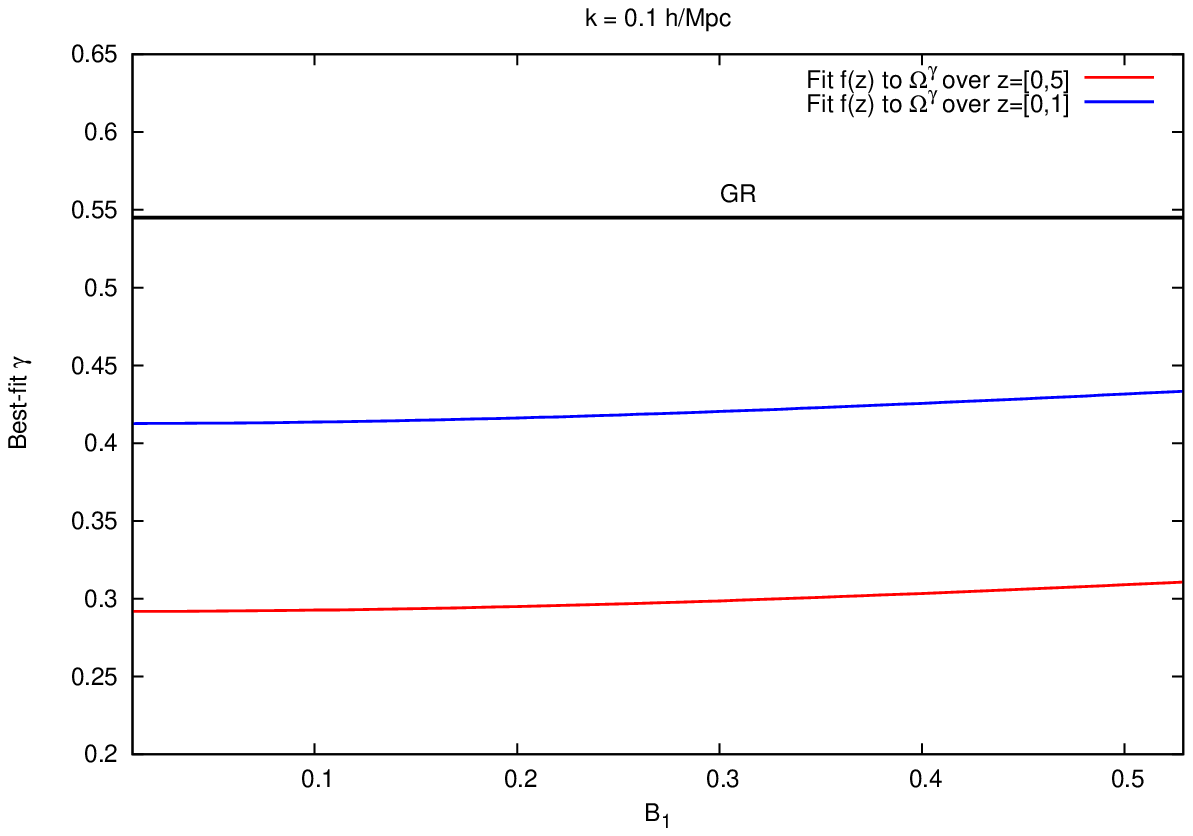}
\includegraphics[width=0.49\textwidth]{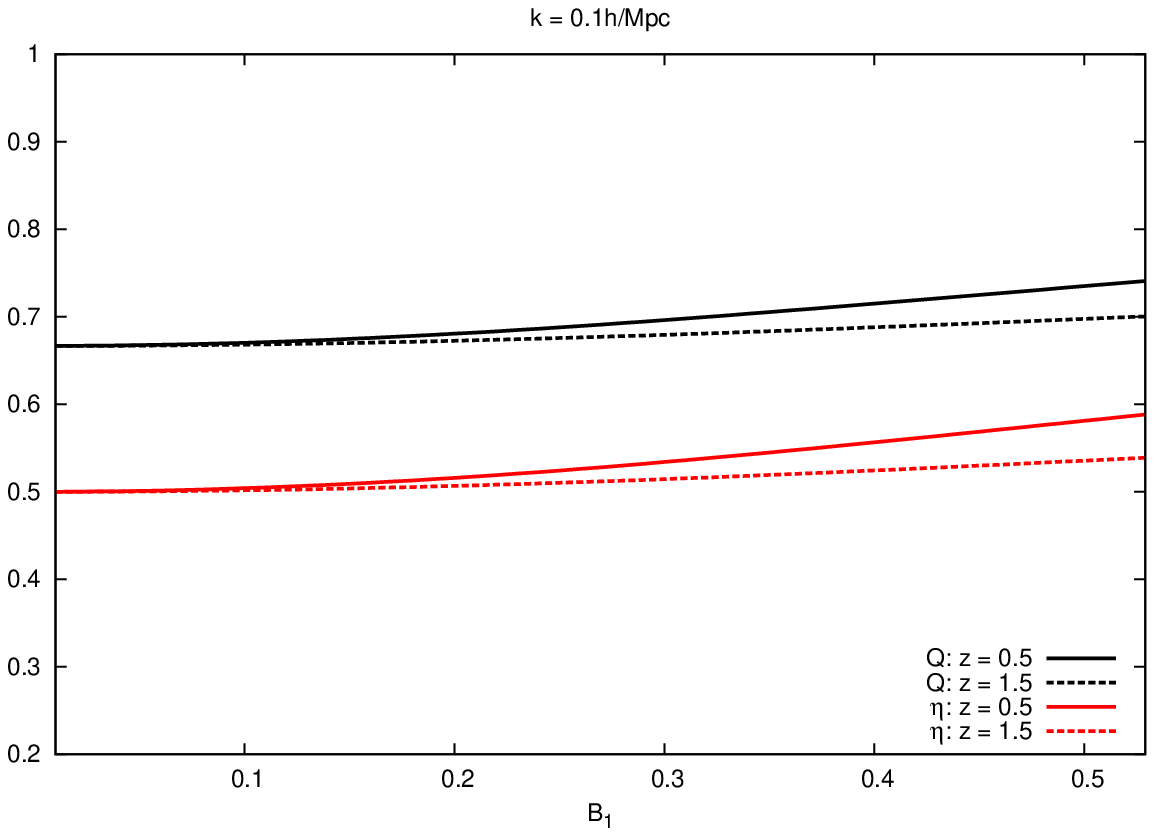}
\includegraphics[width=0.49\textwidth]{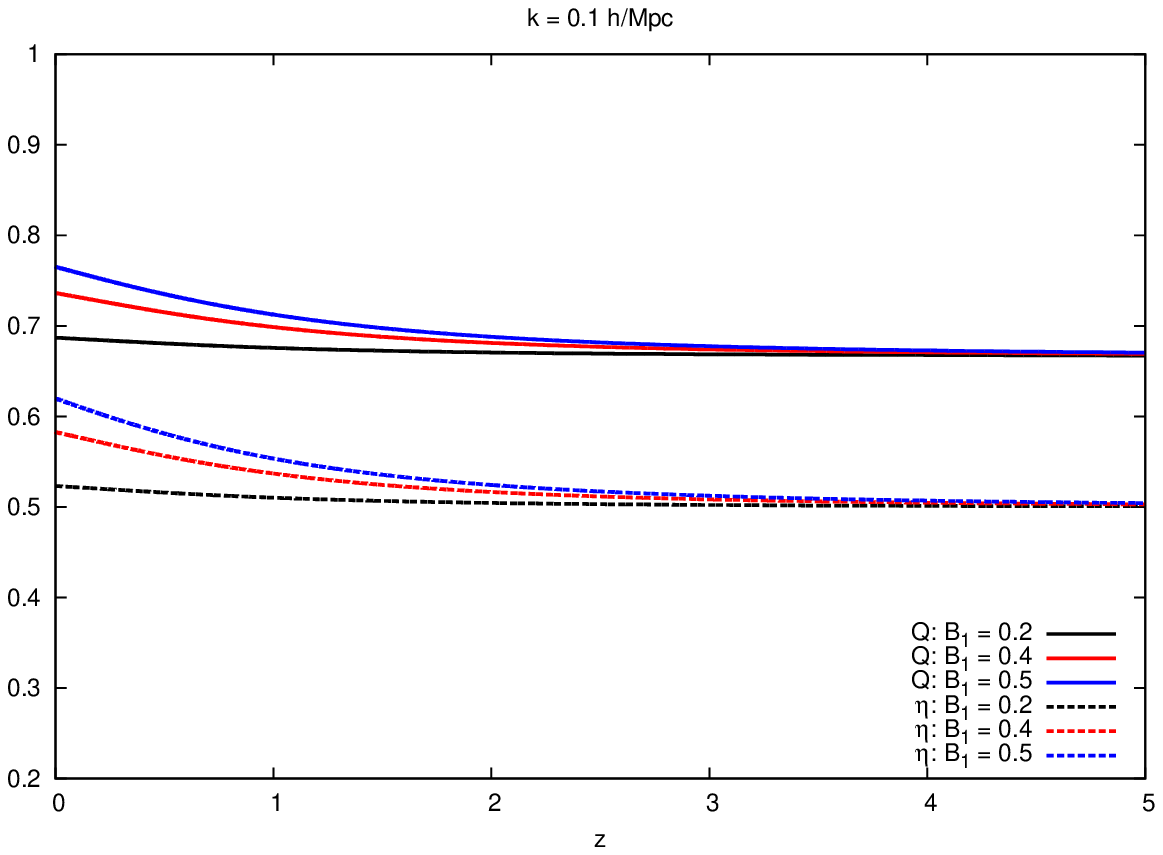}
\caption{Results for the $B_1$--$B_4$ model on the infinite branch, with $\Omega_\Lambda^\mathrm{eff}=0.84$ and $B_1<0.529$. This model is qualitatively different from the others we study, as it does not possess a limit to the minimal $B_1$--only model, and has the strongest deviations from $\Lambda$CDM among all of the models presented in this work. In the first two panels, we plot $f(z)$ as well as the best-fit parametrization $\Omega_m^\gamma$ with the fitting done over the redshift ranges $0<z<1$ and $0<z<5$. In the bottom two panels, we plot $Q$ and $\eta$ and find strong deviations from GR which will be easily visible to near-future LSS experiments.}
\label{fig:b1-b4-infinite}
\end{figure}

Results for the infinite branch are shown in \cref{fig:b1-b4-infinite}. This is the only model we study which does not possess a limit to the minimal $B_1$-only model, and it predicts significant deviations from GR. $\eta$ deviates from 1 by nearly a factor of 2 at all times and for all allowed values of $B_1$, providing a clear observable signal of modified gravity.

$\gamma$ also deviates strongly from the GR value of $\gamma\approx0.545$, and is even lower than the values $\sim0.45$--$0.5$ which we found in the $B_1$--only model and in the other two-parameter models. However, it is worth noting that the $\Omega_m^\gamma$ parametrization is an especially bad fit to $f(z)$ in this case, and another parametrization (see, e.g., section 1.8.3 of Ref. \cite{Amendola:2012ys}) or even a direct comparison between $f(z)$ (multiplied by the growth rate, $G(z)$ \cite{Amendola:2012ys,Macaulay:2013swa}) and the data would be more appropriate. We fit $\gamma$ to $f(z)$ in the redshift range $0<z<5$ (as in the rest of this paper) and $0<z<1$ (which is the redshift range of present observations \cite{Macaulay:2013swa}) in the first and second panels of \cref{fig:b1-b4-infinite}, obtaining significantly different results.

\paragraph{$\mathbf{B_2}$--$\mathbf{B_3}$:}

This is a ``bouncing" model in the $f$ metric: $y$ becomes negative in finite time going towards the past, implying that the $f$ scale factor $Y^2$ collapses to zero and then expands \cite{Konnig:2013gxa}. This itself may not render the model cosmologically unviable. However, we have solved \cref{eq:brepyp} numerically for $y(z)$, and found that $y$ generically goes to $-\infty$ at finite $z$, which means that at higher redshifts there is not a sensible background cosmology. This problem can be avoided by introducing new physics at those higher redshifts to modify the evolution of $y$, or by increasing $B_2$ enough that the pole in $y$ occurs at an unobservably high redshift.\footnote{For $B_2=(5,50,500)$, and $B_3$ chosen to give an effective $\Omega_\Lambda^\mathrm{eff}\approx0.7$ today, the pole occurs at $z\approx(1.99,8.19,27.95)$.} However, these are nonminimal solutions, and so we do not study the $B_2$--$B_3$ model.

\section{Conclusions}
\label{c}

In this paper we have examined the evolution of cosmological perturbations on subhorizon scales, describing linear structure formation during the matter-dominated era, in the massive, ghost-free theory of two interacting metrics. We have studied the most common version of the theory in the literature, in which the matter fields couple to only one of the metrics and couple minimally. We derived the perturbed Einstein equations around FLRW backgrounds and then solved the linearized equations in the subhorizon limit after choosing an appropriate gauge. This gave us the evolution equations for cosmic structure, as well as equations that relate the metric perturbations to the density of matter.

The gauge freedom available to us is the diagonal subgroup of the two diffeomorphism groups for the two metrics; in practice, this means we have a single coordinate chart on which we are free to perform infinitesimal coordinate transformations as in GR, with the same amount of gauge freedom. We chose to work in a Newtonian gauge for the $g$ metric, although our results were not sensitive to this choice (for example, we were able to reproduce our results in the gauge of Ref. \cite{Konnig:2014dna}, $F_g=F_f=0$). In the subhorizon limit, we found that the perturbations were described by a system of six algebraic equations for five variables, and obtained a consistent solution. This solution relates each of the metric perturbations appearing in the subhorizon Einstein equations to the matter density. This allowed us to derive a modified evolution equation for the density contrast, which differs from its GR counterpart by a varying effective Newton's constant, as well as algebraic expressions for the anisotropic stress, $\eta(z,k)$, and the parameter measuring modifications to Newton's constant in the Poisson equation, $Q(z,k)$, purely in terms of background quantities. We then solved for the background numerically to obtain these parameters, and finally integrated the structure growth equation to derive the growth rate, $f(z,k)$ and its best-fit parametrization, $f\sim\Omega_m^\gamma$.

We have studied every subset of the theory which is a) viable at the background level, and b) contains one or two free parameters, while excluding the $g$-metric cosmological constant as we are interested in self-accelerating theories that differ from $\Lambda$CDM. Among the single-parameter models, only the case with the lowest-order interaction term, $\beta_1\neq0$, is in agreement with the background data. We have found that it predicts modified gravity parameters that differ significantly from GR: $\gamma\sim0.46$--$0.48$ (in agreement with Ref. \cite{Konnig:2014dna}), $Q\sim0.94$--$0.95$, and $\eta\sim0.88$--$0.90$. For reference, the $\Lambda$CDM predictions are $\gamma\approx0.545$ and $Q=\eta=1$. Future large-scale structure experiments such as Euclid \cite{Laureijs:2011gra,Amendola:2012ys} will easily be able to distinguish this simple model from GR. As emphasized in Refs. \cite{Akrami:2012vf,Akrami:2013pna,Konnig:2013gxa}, this ``minimal'' bigravity model is especially appealing because it possesses late-time acceleration and fits the background data well with the same number of free parameters as $\Lambda$CDM. It suffers from an early-time instability \cite{Konnig:2014xva}; however, if this can be overcome, or if observations are restricted to late times ($z\lesssim0.5$), our results demonstrate that by going to the level of linear perturbations, this theory can be probed in the near future by multiple observables which deviate significantly from general relativity.

We additionally examined the four two-parameter models that are viable in the background, all of which keep $\beta_1>0$ while turning on a second, higher-order interaction term. Two models, in which either the cubic interaction $\beta_3$ or the $f$-metric cosmological constant $\beta_4$ is nonzero (the latter specifically in the ``finite branch,'' which reduces to the minimal model in the $\beta_4\to0$ limit), have similar behavior. They predict GR-like values for all three modified gravity parameters in the limit where $m^2\beta_1/H_0^2$ is large, becoming indistinguishable from $\Lambda$CDM (given a Euclid-like experiment) for $m^2\beta_1/H_0^2\gtrsim3$. These reduce to the predictions of the minimal model in the limit $m^2\beta_1/H_0^2 \approx 1.45$ (the best-fit value for $\beta_1$ in the minimal model). For lower values of $\beta_1$ (corresponding to positive $\beta_3$ or $\beta_4$) these models predict even more dramatic deviations from GR: $\gamma$ can dip to $0.45$, and $\eta$ at recent times can be as low as $\sim0.75$. Euclid is expected to measure these parameters to within about $0.02$ and $0.1$, respectively \cite{Laureijs:2011gra,Amendola:2012ys,Amendola:2013qna}, and therefore has the potential to break the degeneracies between $\beta_1$ and $\beta_3$ or $\beta_1$ and $\beta_4$ which is present at the level of background observations.

The $\beta_1$--$\beta_2$ model has an instability when $\beta_2$, the coefficient of the quadratic term, is negative. This instability does not necessarily rule out the theory, but it might signal the breakdown of linear perturbation theory, in which case nonlinear studies are required in order to understand the formation of structures. This is different from the early-time instability which is present in all but the infinite-branch $\beta_1$--$\beta_4$ model and does not show up in the subhorizon, quasistatic r\'egime. It is possible that these perturbations at some point become GR-like due to the Vainshtein mechanism. Moreover, because the instability occurs at a characteristic redshift (which depends on the $\beta_i$ parameters), there may be an observable excess of cosmic structure around that redshift. We leave this to future work.

The parameter range of the $\beta_1$--$\beta_2$ model over which linear, subhorizon perturbations are stable, $0<\beta_1\lesssim1.45H_0^2/m^2$ (corresponding to $H_0^2/m^2>\beta_2>0$), is quite small; near the large $\beta_1$ end the predictions recover those of the minimal $\beta_1$-only model, while at the small $\beta_1$ end the perturbations can differ quite significantly from GR, with $\gamma$ as low as $\sim0.35$ and $\eta$ as low as 0.6. However, the exact $\beta_1=0$ limit of this theory is already ruled out by background observations \cite{Akrami:2012vf}, so one should take care when comparing the model to observations in the very low $\beta_1$ region of this parameter space.

Finally, we examined the ``infinite branch'' of the $\beta_1$--$\beta_4$ model, in which the ratio of the $f$ metric scale factor to the $g$ metric scale factor starts at infinity and monotonically decreases to a finite value. In the rest of the models we study, the ratio of the two scale factors starts at zero and then increases; consequently in the $\beta_4\to0$ limit, this theory reduces to pure CDM, rather than to the $\beta_1$-only model. The predictions of the infinite-branch $\beta_1$--$\beta_4$ theory deviate strongly from GR. The model predicts a growth rate $f(z)$ which is not well-parametrized by a $\Omega_m^\gamma$ fit, but has best-fit values of $\gamma$ on the low side ($0.3$--$0.4$, depending on the fitting range). The anisotropic stress $\eta$ is almost always below $0.7$ and can even be as low as $0.5$, a factor of two away from the GR prediction. Across its entire parameter space, this model has the most significantly non-GR values of any we study, and is almost certainly the bigravity model which is viable at the background level that structure observations will be able to rule out first.

Significantly, this infinite-branch $\beta_1$--$\beta_4$ model, or \textit{infinite-branch bigravity}, is stable at all times \cite{Konnig:2014xva}. As the only two-parameter massive bigravity model which both self-accelerates and is always linearly stable around cosmological backgrounds, it is a prime target for observational study.

To summarize, the new ghost-free theory of massive bigravity, in which the graviton is given mass and coupled to a new, metric-like massless spin-2 field, predicts structure formation that deviates significantly from the predictions of general relativity, even while matching the cosmic expansion history as well as $\Lambda$CDM does. These deviations are on the cusp of observability given the impending wave of large-scale structure experiments such as Euclid \cite{Laureijs:2011gra,Amendola:2012ys}. This leaves the exciting prospect of observations in the near future ruling out either the simplest theory of a massless graviton or the simplest theory of a massive one. It is worth noting, however, that by making the free parameters large enough, or perhaps by adding a third or fourth interaction term, these constraints can still in principle be evaded. The question remains how to address these remaining corners of the theory, and further --- in the event that large-scale structure experiments observe deviations from GR --- how to distinguish between certain regions of the bigravity parameter space which have similar predictions. Since the theory is an IR modification of gravity, i.e., its new effects are most significant near the horizon, we expect new observational signatures to exist at superhorizon scales. This may affect the inflationary primordial power spectrum at large scales and/or leave detectable effects on the CMB, such as in the integrated Sachs-Wolfe effect. We leave the investigation of such possible superhorizon signatures, as well as a study of the nonlinear structures in this framework, for future work.

\begin{acknowledgments}
We thank Phil Bull, Jonas Enander, Fawad Hassan, and Edvard M\"ortsell for helpful discussions. We are especially grateful to Luca Amendola and Frank K\"onnig for clarifying discussions during the preparation of an early draft of this paper. A.R.S. is supported by the David Gledhill Research Studentship, Sidney Sussex College, University of Cambridge; and by the Isaac Newton Fund and Studentships, University of Cambridge. Y.A. is supported by the European Research Council (ERC) Starting Grant StG2010-257080. We acknowledge the use of resources from the Norwegian national super-computing facilities, NOTUR.
\end{acknowledgments}

\appendix

\section{Full perturbation equations}
\label{app:perteqs}

Defining the perturbed metrics as\footnote{By leaving the lapse $N$ in the $g$ metric general, we retain the freedom to later work in cosmic or conformal time. There is a further practical benefit: since this choice makes the symmetry between the two metrics manifest, and the action is symmetric between $g$ and $f$ as described in \cref{bigravity}, this means the $f$ field equations can easily be derived from the $g$ equations by judicious use of \texttt{ctrl-f}.}
\begin{align}
ds_g^2 &= -N^2(1+E_g)dt^2 + 2Na\partial_iF_gdt dx^i + a^2\left[(1+A_g)\delta_{ij} + \partial_i\partial_jB_g\right]dx^idx^j, \\
ds_f^2 &= -X^2(1+E_f)dt^2 + 2XY\partial_iF_fdt dx^i + Y^2\left[(1+A_f)\delta_{ij} + \partial_i\partial_jB_f\right]dx^idx^j,
\end{align}
the linearized Einstein equations for the $g$ metric are
\begin{itemize}

\item $0-0$:
\begin{equation}
\frac{3H}{N^2}\left(HE_g - \dot{A}_g\right) + \nabla^2 \left[\frac{A_g}{a^2} + H \left(\frac{2F_g}{Na} - \frac{\dot{B}_g}{N^2}\right) \right] + \frac{m^2}{2}yP\left(3\Delta A + \nabla^2\Delta B\right) = \frac{1}{M_g^2}\delta T{}^0{}_0,
\end{equation}

\item $0-i$:
\begin{equation}
\frac{1}{N^2}\partial_i\left(\dot A_g - HE_g\right) + m^2\frac{P}{x + y}\frac{Y}{N} \partial_i\left(xF_f - yF_g\right) = \frac{1}{M_g^2}\delta T{}^0{}_i,
\end{equation}

\item $i-i$:
\begin{align}
&\frac{1}{N^2}\left[\left(2\dot H + 3H^2 - 2\frac{\dot N}{N}H\right)E_g + H\dot E_g - \ddot{A}_g - 3H\dot{A}_g + \frac{\dot N}{N}\dot A_g \right] + \frac{1}{2}\left(\partial_j^2 + \partial_k^2\right)D_g  \nonumber \\
&+ m^2\left[\frac{1}{2}xP\Delta E + yQ\left(\Delta A+ \frac{1}{2}\left(\partial_j^2 + \partial_k^2\right)\Delta B\right)\right] = \frac{1}{M_g^2}\delta T{}^i{}_i, \label{eq:fullgii} \\
D_g &\equiv \frac{A_g + E_g}{a^2} + \frac{H}{N}\left(\frac{4F_g}{a} - \frac{3\dot{B}_g}{N}\right) + \frac{2\dot{F}_g}{Na} - \frac{1}{N^2}\left(\ddot{B}_g - \frac{\dot N}{N}\dot{B}_g\right),
\end{align}

\item Off-diagonal $i-j$:
\begin{equation}
-\frac{1}{2}\partial^i\partial_j D_g - \frac{m^2}{2}yQ\partial^i\partial_j \Delta B = \frac{1}{M_g^2}\delta T{}^i{}_j,
\end{equation}

\end{itemize}
where $H\equiv\dot{a}/a$ is the usual $g$-metric Hubble parameter (in cosmic or conformal time, depending on $N$), $\partial_j^2 + \partial_k^2$ in the $i-i$ spatial Einstein equation refers to derivatives w.r.t. the other two Cartesian coordinates, i.e., $\nabla^2 - \partial^i \partial_i$ where the $i$ indices are not summed over, and
\begin{align}
P &\equiv \beta_1 + 2\beta_2y + \beta_3y^2, \\
Q &\equiv \beta_1 + \left(x + y\right)\beta_2 + xy\beta_3, \\
x &\equiv X/N, \\
y &\equiv Y/a, \\
\Delta A &\equiv A_f - A_g, \\
\Delta B &\equiv B_f - B_g, \\
\Delta E &\equiv E_f - E_g.
\end{align}

The linearized Einstein equations for the $f$ metric are
\begin{itemize}

\item $0-0$:
\begin{equation}
\frac{3K}{X^2}\left(KE_f - \dot{A}_f\right) + \nabla^2 \left[\frac{A_f}{Y^2} + K \left(\frac{2F_f}{XY} - \frac{\dot{B}_f}{X^2}\right) \right] - \frac{m^2}{2M_\star^2}\frac{P}{y^3}\left(3\Delta A + \nabla^2\Delta B\right) = 0,
\end{equation}

\item $0-i$:
\begin{equation}
\frac{1}{X^2}\partial_i\left(\dot A_f - KE_f\right) + \frac{m^2}{M_\star^2}\frac{P}{y^2}\frac{1}{x + y}\frac{a}{X} \partial_i\left(yF_g - xF_f\right) = 0,
\end{equation}

\item $i-i$:
\begin{align}
&\frac{1}{X^2}\left[\left(2\dot K + 3K^2 - 2\frac{\dot X}{X}K\right)E_f + K\dot E_f - \ddot{A}_f - 3K\dot{A}_f + \frac{\dot X}{X}\dot A_f \right] + \frac{1}{2}\left(\partial_j^2 + \partial_k^2\right)D_f  \nonumber \\
&- \frac{m^2}{M_\star^2}\frac{1}{xy^2}\left[\frac{1}{2}P\Delta E + Q\left(\Delta A+ \frac{1}{2}\left(\partial_j^2 + \partial_k^2\right)\Delta B\right)\right] = 0, \\
D_f &\equiv \frac{A_f + E_f}{Y^2} + \frac{K}{X}\left(\frac{4F_f}{Y} - \frac{3\dot{B}_f}{X}\right) + \frac{2\dot{F}_f}{XY} - \frac{1}{X^2}\left(\ddot{B}_f - \frac{\dot X}{X}\dot{B}_f\right),
\end{align}

\item Off-diagonal $i-j$:
\begin{equation}
-\frac{1}{2}\partial^i\partial_j D_f + \frac{m^2}{2M_\star^2}\frac{Q}{xy^2}\partial^i\partial_j \Delta B = 0,
\end{equation}

\end{itemize}
where $K \equiv\dot{Y}/Y$ is the $f$-metric Hubble parameter.

It is worth mentioning that the $g$ $i-i$ equation, (\ref{eq:fullgii}), is identically zero in GR after taking into account the $0-i$, off-diagonal $i-j$, and $\nu=i$ fluid conservation equations and hence gives no information; in massive (bi)gravity, however, it is crucial, and is manifestly only important when $m\neq0$. In a gauge with $F_g=F_f=0$ it has the simple form
\begin{equation}
m^2\left[P\left(xE_f-yE_g\right)+2yQ\Delta A\right]=0. \label{eq:simpleii}
\end{equation}
This should be compared to \cref{eq:iig-simplified}. Moreover, performing the same steps on the $f$ $i-i$ equation, we arrive again at \cref{eq:simpleii}. Hence both $i-i$ equations carry the same information.

Since these equations are arrived at by a fairly lengthy calculation, for the convenience of future generations of bigravitists we now present some helpful intermediate relations. The metric determinants to linear order are
\begin{align}
\operatorname{det} g &=  -N^2a^6\left(1 + E_g + 3A_g + \nabla^2B_g\right), \\
\operatorname{det} f &=  -X^2Y^6\left(1 + E_f + 3A_f + \nabla^2B_f\right).
\end{align}

The matrix $\mathbbm{X} = \sqrt{g^{-1}f}$ is defined in terms of the two metrics as
\begin{equation}
\mathbbm{X}^\mu{}_\rho \mathbbm{X}^\rho{}_\nu \equiv g^{\mu\rho}f_{\rho\nu}. \label{eq:xdef}
\end{equation}

Its background value is simply
\begin{align}
\mathbbm{X}^0{}_0 &= x, \nonumber \\
\mathbbm{X}^i{}_j &= y \delta^i{}_j.
\end{align}

Using this we can solve \cref{eq:xdef} to first order in perturbations to find
\begin{align}
\mathbbm{X}^0{}_0 &= x\left(1 + \frac{1}{2}\Delta E\right), \nonumber \\
\mathbbm{X}^0{}_i &= \frac{1}{x+y}\frac{Y}{N}\left(y\partial_i F_g - x\partial_i F_f\right), \nonumber \\
\mathbbm{X}^i{}_0 &= \frac{1}{x+y}\frac{X}{a}\left(y\partial^i F_f - x \partial^i F_g\right), \nonumber \\
\mathbbm{X}^i{}_j &= y\left[\left(1+\frac{1}{2}\Delta A\right)\delta^i{}_j + \frac{1}{2}\partial^i\partial_j\Delta B \right].
\end{align}

The trace of this is
\begin{equation}
[\mathbbm{X}] = x\left(1 + \frac{1}{2}\Delta E\right) + y\left[3\left(1+\frac{1}{2}\Delta A\right) + \frac{1}{2}\nabla^2\Delta B\right].
\end{equation}

Similarly we can solve for the matrix $\mathbbm{Y} = \sqrt{f^{-1}g}$, although we do not write its components here as they can be found by simply substituting $(N, a, {}_g)$ with $(X, Y, {}_f)$ and vice versa.\footnote{It may also be calculated explicitly or by using the fact that $\mathbbm{Y}$ is simply the matrix inverse of $\mathbbm{X}$, which can be easily inverted to first order.}

We now need the matrices $\mathbbm{X}^2$ and $\mathbbm{X}^3$ and their traces in order to compute the matrices appearing in the mass terms of the Einstein equations. For $\mathbbm{X}^2$ we find
\begin{align}
(\mathbbm{X}^2)^0{}_0 &= x^2(1 + \Delta E), \nonumber \\
(\mathbbm{X}^2)^0{}_i &= \frac{Y}{N}\left(y\partial_i F_g - x\partial_i F_f\right), \nonumber \\
(\mathbbm{X}^2)^i{}_0 &= \frac{X}{a}\left(y\partial^i F_f - x\partial^iF_g\right), \nonumber \\
(\mathbbm{X}^2)^i{}_j &= y^2 \left[ (1+\Delta A)\delta^i{}_j + \partial^i\partial_j\Delta B\right],
\end{align}
with trace
\begin{equation}
\left[\mathbbm{X}^2\right] = x^2 (1 + E_f - E_g) + y^2 \left[ 3(1 + \Delta A) + \nabla^2\Delta B\right].
\end{equation}

$\mathbbm{X}^3$ is given by
\begin{align}
(\mathbbm{X}^3)^0{}_0 &= x^3\left(1 + \frac{3}{2}\Delta E\right), \nonumber \\
(\mathbbm{X}^3)^0{}_i &= \frac{Y}{N}\left(x + y - \frac{xy}{x+y}\right) \left(y\partial_i F_g - x \partial_i F_f\right), \nonumber \\
(\mathbbm{X}^3)^i{}_0 &= \frac{X}{a}\left(x + y - \frac{xy}{x+y}\right) \left(y\partial^i F_f - x \partial^i F_g\right), \nonumber \\
(\mathbbm{X}^3)^i{}_j &= y^3 \left[\left(1+\frac{3}{2}\Delta A\right)\delta^i{}_j + \frac{3}{2}\partial^i\partial_j\Delta B \right],
\end{align}
with trace
\begin{equation}
\left[\mathbbm{X}^3\right] = x^3\left(1 + \frac{3}{2}\Delta E\right) + y^3 \left[3\left(1+\frac{3}{2}\Delta A\right)+ \frac{3}{2}\nabla^2\Delta B \right].
\end{equation}

$\mathbbm{Y}^2$ and $\mathbbm{Y}^3$ can be determined trivially from these.

With these we can determine the functions $Y^\mu_{(n)\nu}\left(\sqrt{g^{-1}f}\right)$ and $Y^\mu_{(n)\nu}\left(\sqrt{f^{-1}g}\right)$. Two helpful intermediate results are
\begin{align}
\frac{1}{2}\left([\mathbbm{X}]^2 - [\mathbbm{X}^2]\right) &= y^2 \left[3(1 + \Delta A) + \nabla^2 \Delta B\right] \nonumber \\
&\hphantom{{}=} + xy\left[3\left(1+\frac{1}{2}\left(\Delta A + \Delta E\right)\right) + \frac{1}{2}\nabla^2 \Delta B\right], \\
\frac{1}{6}\left([\mathbbm{X}]^3 - 3[\mathbbm{X}][\mathbbm{X}^2] + 2[\mathbbm{X}^3]\right) &= y^3\left(1 + \frac{3}{2}\Delta A + \frac{1}{2}\nabla^2\Delta B\right) \nonumber \\
&\hphantom{{}=} + xy^2\left(3\left(1 + \Delta A + \frac{1}{2}\Delta E\right) + \nabla^2\Delta B\right).
\end{align}

To obtain those intermediate results and the $0-0$ and $i-j$ components of the $Y$ matrices, it saves a lot of algebra to write the traces, $0-0$ components, and $i-j$ components of the various $\mathbbm{X}$ matrices in terms of
\begin{align}
c_1 &= x, & c_2 &= 3y, \nonumber \\
 \delta_1 &= \frac{1}{2}\Delta E, & \delta_2 &= \frac{1}{2}\Delta A + \frac{1}{6}\nabla^2\Delta B, && \delta_3{}^i{}_j = \left(\frac{1}{2}\partial^i\partial_j - \frac{1}{6}\delta^i{}_j\nabla^2\right)\Delta B.
\end{align}

Finally, the matrices $Y^\mu_{(n)\nu}\left(\sqrt{g^{-1}f}\right)$ defined in \cref{eq:ymatdef} are given by:
\begin{itemize}

\item $n=0$:
\begin{equation}
Y^\mu_{(0)\nu}\left(\sqrt{g^{-1}f}\right) = \delta^\mu{}_\nu,
\end{equation}

\item $n=1$:
\begin{align}
Y^0_{(1)0}\left(\sqrt{g^{-1}f}\right) &= -y\left[3\left(1+\frac{1}{2}\Delta A\right) + \frac{1}{2}\nabla^2\Delta B\right], \nonumber \\
Y^0_{(1)i}\left(\sqrt{g^{-1}f}\right) &= \frac{1}{x+y}\frac{Y}{N}\left(y\partial_i F_g - x \partial_i F_f\right), \nonumber \\
Y^i_{(1)0}\left(\sqrt{g^{-1}f}\right) &= \frac{1}{x+y}\frac{X}{a}\left(y\partial^i F_f - x \partial^i F_g\right), \nonumber \\
Y^i_{(1)j}\left(\sqrt{g^{-1}f}\right) &= -x\left(1 + \frac{1}{2}\Delta E\right)\delta^i{}_j \nonumber\\
&\hphantom{{}=} - 2y\left[\left(1+\frac{1}{2}\Delta A\right)\delta^i{}_j + \frac{1}{4}\left(\delta^i{}_j\nabla^2 - \partial^i\partial_j\right)\Delta B\right],
\end{align}

\item $n=2$:
\begin{align}
Y^0_{(2)0}\left(\sqrt{g^{-1}f}\right) &= y^2 \left[ 3(1 + \Delta A) + \nabla^2\Delta B\right], \nonumber \\
Y^0_{(2)i}\left(\sqrt{g^{-1}f}\right) &= -\frac{2y}{x+y} \frac{Y}{N}\left(y\partial_i F_g - x \partial_i F_f\right), \nonumber \\
Y^i_{(2)0}\left(\sqrt{g^{-1}f}\right) &= -\frac{2y}{x+y} \frac{X}{a}\left(y\partial^i F_f - x \partial^i F_g\right), \nonumber \\
Y^i_{(2)j}\left(\sqrt{g^{-1}f}\right) &= y^2\left[(1 + \Delta A)\delta^i{}_j + \frac{1}{2}\left(\delta^i{}_j\nabla^2 - \partial^i\partial_j\right)\Delta B\right] \nonumber \\
&\hphantom{{}=} + 2xy\left[ \left(1 + \frac{1}{2}(\Delta A + \Delta E)\right)\delta^i{}_j + \frac{1}{4} \left(\delta^i{}_j \nabla^2 - \partial^i\partial_j\right)\Delta B \right],
\end{align}

\item $n=3$:
\begin{align}
Y^0_{(3)0}\left(\sqrt{g^{-1}f}\right) &= -y^3\left[1 + \frac{3}{2}\Delta A + \frac{1}{2}\nabla^2\Delta B\right], \nonumber \\
Y^0_{(3)i}\left(\sqrt{g^{-1}f}\right) &= \frac{y^2}{x + y} \frac{Y}{N} \left(y\partial_i F_g - x \partial_i F_f\right), \nonumber \\
Y^i_{(3)0}\left(\sqrt{g^{-1}f}\right) &= \frac{y^2}{x+y} \frac{X}{a}\left(y\partial^i F_f - x \partial^i F_g\right), \nonumber \\
Y^i_{(3)j}\left(\sqrt{g^{-1}f}\right) &= -xy^2\left[\left(1 + \Delta A + \frac{1}{2}\Delta E\right)\delta^i{}_j + \frac{1}{2}\left(\delta^i{}_j\nabla^2 - \partial^i\partial_j\right)\Delta B\right].
\end{align}

\end{itemize}

Plugging these matrices into the field \cref{eq:Einsteingeng,eq:Einsteingenf}, we obtain the full perturbation equations presented at the beginning of this appendix.

\section{Explicit solutions for the modified gravity parameters}
\label{app:hcoeff}

As discussed in \cref{results}, the modified gravity parameters $Q$ and $\eta$ have the Horndeski form,
\begin{align}
\eta &= h_2\left(\frac{1 + k^2h_4}{1+k^2h_5}\right), \\
Q &= h_1\left(\frac{1 + k^2h_4}{1+k^2h_3}\right),
\end{align}
while the growth of structures can be written in these terms as
\begin{equation}
\ddot \delta + H\dot\delta - \frac{1}{2}\frac{Q}{\eta}\frac{a^2\bar\rho}{M_g^2}\delta = 0.
\end{equation}

Hence the five $h_i$ coefficients allow us to determine all three modified gravity parameters we consider in this paper. They are given explicitly by
\begin{align}
h_1 &= \frac{1}{1+y^2}, \\
h_2 &= -\frac{\left(1+y^2\right) \left(\beta _1+3 \beta _3 y^4+\left(6 \beta _2-2 \beta _4\right) y^3+3 \left(\beta _1-\beta _3\right) y^2\right)}{-\beta   _1+\left(3 \beta _2-\beta _4\right) y^5+\left(6 \beta _1-9 \beta _3\right) y^4+\left(3 \beta _0-15 \beta _2+2 \beta _4\right) y^3+\left(3 \beta _3-7   \beta _1\right) y^2}, \\
h_3 &= -\frac{y^2}{h_6}\frac{3}{1+y^2}\bigg[\beta _3^2 y^7+\left(4 \beta _2 \beta _3-2 \beta _3 \beta _4\right) y^6+3 \left(2 \beta _1-3 \beta_3\right) \beta _3 y^5 +\left(4 \beta _0 \beta _3-19 \beta _2 \beta _3-\beta _4 \beta _3+2 \beta _1 \beta _4\right) y^4  \nonumber \\
   &\hphantom{{}=-y^2\bigg[} +\left(-3 \beta _1^2-18 \beta_2^2-6 \beta _3^2+4 \beta _0 \beta _2+2 \beta _2 \beta _4\right) y^3 +3 \left(\left(\beta _0-3 \beta _2\right) \beta _3+\beta _1 \left(\beta _4-5 \beta_2\right)\right) y^2  \nonumber \\
   &\hphantom{{}=-y^2\bigg[} +\left(-7 \beta _1^2+2 \beta _3 \beta _1+2 \left(\beta _0-3 \beta _2\right) \beta _2\right) y -\beta _1 \left(\beta _0+\beta _2\right)\bigg], \\
h_4 &= \frac{y^2}{h_6}\bigg[2 \beta _3 \beta _4 y^6+2 \left(3 \beta _2^2-\beta _4 \beta _2-3 \left(\beta _1-2 \beta _3\right)
   \beta _3\right) y^5 +\left(\beta _1 \left(6 \beta _2-4 \beta _4\right)+3 \beta _3 \left(-2 \beta _0+9 \beta _2+\beta _4\right)\right) y^4 \nonumber\\
   &\hphantom{{}=y^2\bigg[}+2 \left(3 \beta
   _1^2-2 \beta _3 \beta _1+18 \beta _2^2+9 \beta _3^2-3 \beta _0 \beta _2-3 \beta _2 \beta _4\right) y^3 +\left(37 \beta _1 \beta _2+27 \beta _3 \beta _2-9
   \beta _0 \beta _3-9 \beta _1 \beta _4\right) y^2 \nonumber\\
   &\hphantom{{}=y^2\bigg[}+2 \left(10 \beta _1^2-3 \beta _3 \beta _1-3 \left(\beta _0-3 \beta _2\right) \beta _2\right) y + 3 \beta _1 \left(\beta _0+\beta _2\right)\bigg], \\ 
h_5 &= -\frac{y^2}{h_6}\frac{1}{h_7}\bigg[4 \beta _3^2 \beta _4^2 y^{11}+\beta _3 \left(24 \beta _4 \beta _2^2+\left(9 \beta _3^2-8 \beta _4^2\right) \beta _2+3 \beta _3 \left(19 \beta _3-8 \beta
   _1\right) \beta _4\right) y^{10} \nonumber\\
   &\hphantom{{}=-\frac{y^2}{h_6}\frac{1}{h_7}\bigg[} +2 \left(18 \beta _2^4-12 \beta _4 \beta _2^3+\left(117 \beta _3^2-36 \beta _1 \beta _3+2 \beta _4^2\right) \beta _2^2+6
   \beta _3 \left(4 \beta _1+5 \beta _3\right) \beta _4 \beta _2 \right. \nonumber\\
   &\hphantom{{}=-\frac{y^2}{h_6}\frac{1}{h_7}\bigg[+2(} \left.+\beta _3 \left(99 \beta _3^3-81 \beta _1 \beta _3^2+18 \beta _1^2 \beta _3-3 \beta _4^2
   \beta _3-12 \beta _0 \beta _4 \beta _3-8 \beta _1 \beta _4^2\right)\right) y^9 \nonumber\\
   &\hphantom{{}=-\frac{y^2}{h_6}\frac{1}{h_7}\bigg[}-\left(72 \beta _3 \left(\beta _2-\beta _4\right) \beta _1^2+\left(-72
   \beta _2^3+72 \beta _4 \beta _2^2+\left(117 \beta _3^2-16 \beta _4^2\right) \beta _2+\beta _3^2 \left(85 \beta _4-72 \beta _0\right)\right) \beta _1 \right. \nonumber\\
   &\hphantom{{}=-\frac{y^2}{h_6}\frac{1}{h_7}\bigg[-(} \left.+9
   \beta _3 \left(-60 \beta _2^3+8 \beta _0 \beta _2^2+12 \beta _4 \beta _2^2-96 \beta _3^2 \beta _2+19 \beta _0 \beta _3^2+5 \beta _3^2 \beta
   _4\right)\right) y^8 \nonumber\\
   &\hphantom{{}=-\frac{y^2}{h_6}\frac{1}{h_7}\bigg[}-2 \left(36 \beta _3 \beta _1^3-\left(54 \beta _2^2-36 \beta _4 \beta _2+69 \beta _3^2+8 \beta _4^2\right) \beta _1^2 \right. \nonumber\\
   &\hphantom{{}=-\frac{y^2}{h_6}\frac{1}{h_7}\bigg[-2(} \left.-2 \beta _3
   \left(123 \beta _2^2-76 \beta _4 \beta _2+3 \left(13 \beta _3^2+\beta _4 \left(4 \beta _0+\beta _4\right)\right)\right) \beta _1 \right.\nonumber\\
   &\hphantom{{}=-\frac{y^2}{h_6}\frac{1}{h_7}\bigg[-2(} \left.-3 \left(72 \beta _2^4-36
   \beta _4 \beta _2^3+\left(255 \beta _3^2+4 \beta _4^2\right) \beta _2^2-21 \beta _3^2 \beta _4 \beta _2-9 \beta _3^4+6 \beta _0^2 \beta _3^2 \right.\right. \nonumber\\
   &\hphantom{{}=-\frac{y^2}{h_6}\frac{1}{h_7}\bigg[-2(-3(} \left.\left.+\beta _0
   \left(-12 \beta _2^3+4 \beta _4 \beta _2^2-93 \beta _3^2 \beta _2+3 \beta _3^2 \beta _4\right)\right)\right) y^7 \nonumber\\
   &\hphantom{{}=-\frac{y^2}{h_6}\frac{1}{h_7}\bigg[}+\left(24 \left(3 \beta _2-2 \beta
   _4\right) \beta _1^3+\beta _3 \left(-72 \beta _0+507 \beta _2-77 \beta _4\right) \beta _1^2 \right. \nonumber\\
   &\hphantom{{}=-\frac{y^2}{h_6}\frac{1}{h_7}\bigg[+(}\left.+\left(876 \beta _2^3-508 \beta _4 \beta _2^2+600 \beta _3^2
   \beta _2+48 \beta _4^2 \beta _2-3 \beta _3^2 \beta _4+\beta _0 \left(-72 \beta _2^2+48 \beta _4 \beta _2-69 \beta _3^2\right)\right) \beta _1 \right. \nonumber\\
      &\hphantom{{}=-\frac{y^2}{h_6}\frac{1}{h_7}\bigg[+(} \left. +3 \beta _3
   \left(24 \beta _2 \beta _0^2+\left(-228 \beta _2^2+16 \beta _4 \beta _2+9 \beta _3^2\right) \beta _0+9 \beta _2 \left(48 \beta _2^2-4 \beta _4 \beta _2-7
   \beta _3^2\right)\right)\right) y^6 \nonumber\\
      &\hphantom{{}=-\frac{y^2}{h_6}\frac{1}{h_7}\bigg[}+2 \left(18 \beta _1^4+45 \beta _3 \beta _1^3+\left(477 \beta _2^2-36 \beta _0 \beta _2-170 \beta _4 \beta _2+14 \beta
   _3^2+9 \beta _4^2\right) \beta _1^2 \right.\nonumber\\
      &\hphantom{{}=-\frac{y^2}{h_6}\frac{1}{h_7}\bigg[+2(} \left.+3 \beta _3 \left(126 \beta _2^2-42 \beta _0 \beta _2+6 \beta _4 \beta _2-3 \beta _3^2+2 \beta _0 \beta _4\right)
   \beta _1 \right. \nonumber\\
      &\hphantom{{}=-\frac{y^2}{h_6}\frac{1}{h_7}\bigg[+2(} \left.+6 \left(\beta _0-3 \beta _2\right) \beta _2 \left(-15 \beta _2^2+3 \beta _0 \beta _2+2 \beta _4 \beta _2+6 \beta _3^2\right)\right)
   y^5 \nonumber \\
      &\hphantom{{}=-\frac{y^2}{h_6}\frac{1}{h_7}\bigg[} +\left(\left(441 \beta _2-79 \beta _4\right) \beta _1^3+\beta _3 \left(-33 \beta _0-8 \beta _2+33 \beta _4\right) \beta _1^2 \right.\nonumber\\
         &\hphantom{{}=-\frac{y^2}{h_6}\frac{1}{h_7}\bigg[+(} \left.+\left(648 \beta _2^3-156
   \beta _0 \beta _2^2-60 \beta _4 \beta _2^2+9 \beta _3^2 \beta _2+9 \beta _0 \beta _3^2\right) \beta _1+36 \left(\beta _0-3 \beta _2\right) \beta _2^2
   \beta _3\right) y^4 \nonumber \\
      &\hphantom{{}=-\frac{y^2}{h_6}\frac{1}{h_7}\bigg[}+2 \beta _1 \left(39 \beta _1^3-26 \beta _3 \beta _1^2+\left(167 \beta _2^2-15 \beta _4 \beta _2+15 \beta _3^2-3 \beta _0 \left(\beta
   _2+\beta _4\right)\right) \beta _1-12 \beta _0 \beta _2 \beta _3\right) y^3 \nonumber\\
   &\hphantom{{}=-\frac{y^2}{h_6}\frac{1}{h_7}\bigg[}+\beta _1 \left(36 \beta _2^3+9 \beta _1 \beta _3 \beta _2+3 \beta _0 \left(3
   \beta _1^2-5 \beta _3 \beta _1-4 \beta _2^2\right)+\beta _1^2 \left(112 \beta _2-9 \beta _4\right)\right) y^2 \nonumber\\
   &\hphantom{{}=-\frac{y^2}{h_6}\frac{1}{h_7}\bigg[}+2 \beta _1^2 \left(11 \beta _1^2-3 \beta _3
   \beta _1+12 \beta _2^2\right) y+3 \beta _1^3 \left(\beta _0+\beta _2\right)\bigg], \\
h_6 &=  3 m^2a^2 \left(1+y^2\right) \left(\beta _1+\beta _3 y^2+2 \beta _2 y\right) \left(\beta _1^2+\beta _3 \beta _4 y^5+\left(3 \beta _2^2-\beta _4 \beta _2-3
   \left(\beta _1-2 \beta _3\right) \beta _3\right) y^4 \right.\nonumber\\
   &\hphantom{{}=\bigg[} \left. +\left(3 \beta _1 \beta _2+12 \beta _3 \beta _2-3 \beta _0 \beta _3-2 \beta _1 \beta _4\right)
   y^3+\left(3 \beta _1^2+\beta _3 \beta _1-3 \left(\beta _0-3 \beta _2\right) \beta _2\right) y^2+5 \beta _1 \beta _2 y\right), \\
h_7 &= \frac{\left(\beta _1+y \left(2 \beta _2+\beta _3 y\right)\right) \left(3 \beta _0 y^3+y^2 \left(3 \beta _2 y \left(y^2-5\right)+\beta _3 \left(3-9
   y^2\right)-\beta _4 y \left(y^2-2\right)\right)+\beta _1 \left(6 y^4-7 y^2-1\right)\right)}{1+y^2}.
\end{align}
While this notation is inspired by Refs. \cite{Amendola:2012ky,Amendola:2013qna}, we have defined $h_{1,6,7}$ differently.

\bibliographystyle{JHEP}
\bibliography{bibliography}

\end{document}